# Task-Aware Identifiability: Observation Quotients, Statistical Geometry, and Representation Accessibility

Yan Jiao[1], Pin-Han Ho[1], Limei Peng[1,2], Yiming Miao[1]

[1] Shenzhen Institute for Advanced Study,
University of Electronic Science and Technology of China;
[2] Kyungpook National University
Correspondence: Pin-Han Ho, pinhanho71@gmail.com

## Abstract

Intelligent systems often assume observations contain the information required for a task. This fails structurally when a declared mechanism assigns identical observation laws to latent states the task must distinguish. We develop a unified diagnostic framework connecting observation-induced equivalence, task identifiability, local statistical geometry, representation accessibility, and achievable risk. Its organizing condition is a standard factorization property: the observation partition refines the task partition, and conditionally iid repetitions of that fixed channel cannot recover an exactly collapsed distinction. On regular quotient strata, Fisher geometry and nuisance-efficient Fisher information characterize first-order local resolvability. For representations, we distinguish law-valued encoder experiments from deterministic quotient embeddings and exact task sufficiency from scale-controlled decoder accessibility. We organize task risk into structural, finite-observation, representation-channel, and learning gaps. Controlled synthetic studies validate these layers; a physics-based synthetic near-field model yields sharply different range and angle information, while NYUv2 shows persistent, decoder-modulated task preferences among frozen encoders under two pointwise probes. TAI thereby localizes whether task-relevant information is absent at observation, weakly resolved, lost or poorly accessible in representation, or left to the learner.

# 1 Introduction

## 1.1 The Problem of Knowability

Intelligent systems often assume that their observations contain sufficient information to recover task-relevant states. Under standard formulations such as $p(y \mid x)$, learning then focuses on estimating the relation between observations and states.

However, a fundamental question precedes learning: **Does the observation process itself preserve the distinctions the agent needs to make?**

Consider an idealized monocular observation model in which texture, motion, known scale, and other side information are absent. Under this simplified projection, distinct corridor depths may map to the same image—not because the neural network is too small or the training set is insufficient, but because the declared observation function collapses those depths. This illustrates exact structural loss under that model, not a universal claim about monocular perception; such loss is distinct from—though able to coexist with—graded statistical uncertainty.

Identifiability is not fundamentally a property of a learner; it is a compatibility property between the equivalence relation induced by observation and the distinctions required by a task. This paper formalizes this compatibility through the observation quotient and its geometric consequences.

## 1.2 What the Theory Provides

For orientation, a latent state may be written as $u = (x, v) \in U = d' \times N$, with task-relevant variable $x$ and nuisance variable $v$. A declared observation mechanism assigns the conditional law $P_{Y|u}$ to each state. Two states are observationally equivalent when

$$u_1 \sim_O u_2 \quad \Longleftrightarrow \quad P_{Y|u_1} = P_{Y|u_2}, \tag{1}$$

with equality as probability measures. The observation quotient

$$Q_O = U / \sim_O \tag{2}$$

collects the distinctions preserved by the declared mechanism. For a downstream task $h : U \to H$, let $u_1 \sim_h u_2$ when $h(u_1) = h(u_2)$. The task-aware compatibility condition is

$$\boxed{\sim_O \in \sim_h}. \tag{3}$$

Equivalently, the observation partition refines the task partition. Compatibility gives structural task identifiability; failure leaves an ambiguity no observation-only procedure can resolve. Section 3 develops these statements formally. Subsequent sections examine local statistical strength, representation accessibility, and the resulting layers of task risk.

## 1.3 Contributions

This paper develops a unified task-aware identifiability framework connecting observation-induced equivalence,

structural task identifiability, local statistical geometry, representation accessibility, and achievable task performance. Specifically:

1. **Structural observation quotient and task identifiability**: We organize standard quotient-factorization and decision-theoretic ideas into a task-aware diagnostic: structural identifiability holds iff the observation partition refines the task partition, a condition invariant under conditionally iid repetitions of the same fixed channel.

2. **First-order local identifiability via Fisher geometry**: On regular quotient strata, standard Fisher geometry induces a Riemannian metric via pullback from the statistical image. Its nullspace characterizes first-order locally insensitive directions without being conflated with global observational equivalence.

3. **Task-relative representation accessibility**: We distinguish the law-valued experiment induced by an actual encoder from an idealized deterministic quotient embedding. Exact fixed-task sufficiency, a scale-controlled task-decoder Lipschitz modulus, and local task pullback geometry separate information preservation from geometric accessibility; global quotient distortion remains a strong task-agnostic sufficient bound.

4. **Layered risk and controlled diagnostics**: We separate structural, finite-observation, representation-channel, and learning gaps. Validation spans controlled synthetic studies, a physics-based synthetic ULA sensing study, and an NYUv2 evaluation of frozen learned representations across the framework's diagnostic levels.

# 2 Related Work

Before developing the framework, we position TAI relative to the main literatures that supply its component ideas: statistical identifiability and observability, comparison of statistical experiments, information geometry, representation learning, and active perception. The goal is not to claim novelty for these individual ingredients, but to clarify the diagnostic role assigned to each within the task-aware framework developed in the following sections.

## 2.1 Statistical Identifiability and Observability

Classical identifiability and observability characterize when parameters or states can be recovered [1, 2, 4, 5, 6]. Modern analyses further distinguish structural non-identifiability from finite- or noise-induced practical non-identifiability [24]. TAI organizes these questions around task compatibility with the observation-induced equivalence relation and connects the resulting boundary to representation diagnostics.

## 2.2 Sufficient Statistics and Blackwell Ordering

Sufficient statistics compress observations while preserving parameter information. Blackwell comparison orders statistical experiments through decision-theoretic informativeness and garbling [3]. Broader comparison-of-experiments theory treats randomization, deficiency, and risk comparison [25, 26]. Information Bottleneck and sufficient-representation work formalize retention of task-relevant information [17, 18]; predictive $\mathcal{V}$-information and the Decodable Information Bottleneck additionally make decoder-family restrictions explicit [19, 20]. TAI uses these established foundations while exposing the equivalence structure induced on the latent domain by a fixed experiment and separating exact fixed-task sufficiency from decoder conditioning.

## 2.3 Information Geometry

Information geometry studies statistical manifolds and Fisher–Rao structure [7, 8]. Fisher monotonicity under sufficient statistics and Markov maps is standard [21]. TAI uses the Fisher nullspace as a local, first-order probe of the observation quotient and keeps this law-level geometry distinct from deterministic representation pullbacks.

## 2.4 State Abstractions and Representation Learning

State abstractions aggregate equivalent states [12, 13, 27]. Identifiable representation learning aims to recover latent variables [9, 10, 11], while stable embedding theory studies metric preservation and inverse stability [22]. Task-relevant latent identification has also been studied under structured invertible generative models [28], where the target is recovery of task-relevant latent representations. Empirical transfer studies also show that representation usefulness depends on the target task [23]. Contrastive-learning analyses likewise show that the information retained by a representation can depend on the downstream task [29, 30]. Related representation-evaluation work has also used risk decomposition to separate representation usability from encoder and probe generalization [31]. TAI's complementary contribution is the joint diagnostic separation of observation availability, fixed-task sufficiency, scale-controlled decoder stability, and representation-channel risk.

## 2.5 Active Perception

Active perception [14] and Bayesian experimental design [15, 16] select informative actions. Task-based quantization also designs acquisition front ends for specified downstream tasks [32]. The present paper does not define an action-dependent sensing objective; its task-conditioned diagnostics may motivate such formulations in future work.

# 3 Observation-Induced Structural Identifiability

## 3.1 Observation Mechanisms

Let $(U, \mathcal{A}_U)$ and $(Y, \mathcal{A}_Y)$ be measurable spaces. An observation mechanism is a Markov kernel

$$O : U \times \mathcal{A}_Y \to [0, 1], \tag{4}$$

where $O(u, A) = P(Y \in A \mid U = u)$.

For deterministic observations, there exists a measurable function $f : U \to Y$ such that

$$O(u, A) = \mathbf{1}\{f(u) \in A\}. \tag{5}$$

**Definition 1** (Statistical Observation Equivalence). *For an observation mechanism $O$, define*

$$u_1 \sim_O u_2 \iff O(u_1, \cdot) = O(u_2, \cdot), \tag{6}$$

*where equality is equality of probability measures on $(Y, \mathcal{A}_Y)$.*

**Definition 2** (Observation Quotient and Partition). *The observation quotient is*

$$Q_O = U/ \sim_O . \tag{7}$$

*The canonical projection is*

$$\pi_O : U \to Q_O, \qquad \pi_O(u) = [u]_O. \tag{8}$$

*The induced partition is*

$$P_O = \{[u]_O : u \in U\}. \tag{9}$$

**Remark 3.** *For deterministic observations,*

$$u_1 \sim_O u_2 \iff \delta_{f(u_1)} = \delta_{f(u_2)}. \tag{10}$$

*A measurable space is* point-separating *if for every $y_1 \neq y_2$ there is an $A \in \mathcal{A}_Y$ containing exactly one of $y_1, y_2$. If $(Y, \mathcal{A}_Y)$ is point-separating, then equality of the Dirac laws above is equivalent to $f(u_1) = f(u_2)$, and $Q_f = U/ \sim_f$ is the coimage of $f$. Standard Borel observation spaces are point-separating.*

## 3.2 Quotient Regularity

In general, $Q_O$ is a quotient set or measurable quotient. It need not admit a smooth structure. We distinguish three levels:

1. **Measurable quotient**: $Q_O$ with the quotient $\sigma$-algebra.
2. **Regular quotient stratum**: A subset $Q_O^{\mathrm{reg}}$ admitting a smooth manifold structure.
3. **Riemannian quotient**: $Q_O^{\mathrm{reg}}$ endowed with a metric pulled back from the Fisher-Rao geometry.

**Definition 4** (Measurable Quotient Structure). *Equip $Q_O$ with the quotient $\sigma$-algebra*

$$\mathcal{A}_Q = \left\{ B \subseteq Q_O : \pi_O^{-1}(B) \in \mathcal{A}_U \right\}. \tag{11}$$

*Then $\pi_O$ is measurable. Every measurable function $f$ on $U$ that is constant on observation-equivalence classes descends measurably to $Q_O$: for its descended map $\bar{f}$, $\pi_O^{-1}(\bar{f}^{-1}(A)) = f^{-1}(A)$.*

## 3.3 Task-Observation Compatibility

**Definition 5** (Task Equivalence). *For a task $h : U \to H$, define*

$$u_1 \sim_h u_2 \iff h(u_1) = h(u_2). \tag{12}$$

*The task partition is*

$$P_h = U/ \sim_h . \tag{13}$$

**Definition 6** (Refinement Order). *For two partitions $P_1, P_2$ of $U$, we write $P_1 \succeq P_2$ if every cell of $P_1$ is contained in some cell of $P_2$ (i.e., $P_1$ is at least as fine as $P_2$).*

**Theorem 7** (Task-Observation Compatibility). *Let $h : U \to H$ be a task and $O$ an observation mechanism. The following are equivalent:*

1. *There exists $\bar{h} : Q_O \to H$ such that $h = \bar{h} \circ \pi_O$.*
2. *For all $u_1, u_2 \in U$, $u_1 \sim_O u_2 \implies h(u_1) = h(u_2)$.*
3. *$\sim_O \subseteq \sim_h$.*
4. *$P_O \succeq P_h$.*

*Proof.* If $h = \bar{h} \circ \pi_O$, then $u_1 \sim_O u_2$ implies $\pi_O(u_1) = \pi_O(u_2)$, hence $h(u_1) = h(u_2)$. Conversely, if $h$ is constant on observation equivalence classes, define $\bar{h}([u]_O) = h(u)$; this is well-defined by the fiber-constant condition. The equivalence with the refinement and subset conditions follows from the definitions. □

**Remark 8.** *The quotient factorization property itself is standard. The contribution here is its interpretation as the structural condition for task identifiability from observations.*

**Corollary 9** (Structural Identifiability Boundary). *Let $D_h = \{(u_1, u_2) \in U \times U : h(u_1) \neq h(u_2)\}$ be the set of task-distinguishing pairs. A task $h$ is structurally identifiable from $O$ if and only if*

$$D_h \cap (\sim_O) = \emptyset. \tag{14}$$

*When this condition fails, there exists at least one pair $(u_1, u_2)$ with $u_1 \sim_O u_2$ but $h(u_1) \neq h(u_2)$; no observation-only procedure can resolve this structural distinction.*

*Proof.* The condition $D_h \cap (\sim_O) = \emptyset$ says exactly that $u_1 \sim_O u_2$ implies $h(u_1) = h(u_2)$, which is the compatibility condition in Theorem 7. If an equivalent pair has distinct deterministic task labels, every observation-only decision rule has the same output law under the two states and therefore cannot be almost surely correct for both labels. □

## 3.4 Structural Ambiguity vs. Structural Risk

For this subsection, assume that $h$ is measurable and that the relevant regular conditional distributions exist,

and let $P_{U|Q=q}$ denote the regular conditional distribution of $U$ given $Q = \pi_O(U) = q$. The conditional task distribution is

$$P(h(U) \in A \mid Q = q) = \int_{\mathcal{U}} \mathbf{1}\{h(u) \in A\} P_{U|Q=q}(du). \tag{15}$$

For the following 0-1 loss formulas, assume $\mathcal{H}$ is finite, or more generally that the displayed conditional maximum and a measurable maximizing decision exist. The optimal structural quotient risk is

$$R^*_Q = \mathbb{E}_Q \left[ 1 - \max_{y \in \mathcal{H}} P(h(U) = y \mid Q) \right]. \tag{16}$$

Since $h$ is measurable and deterministic, the rule $r = h$ is admissible, and under 0-1 loss $R^*_U = 0$. Thus $\epsilon_{\text{struct}} = R^*_Q$ in this setting.

**Definition 10** (Structural Ambiguity). *The worst-case structural ambiguity is*

$$\Delta_h(\mathcal{O}) = \sup_{u_1 \sim_O u_2} d_{\mathcal{H}}(h(u_1), h(u_2)). \tag{17}$$

Structural ambiguity is distribution-free; structural risk is distribution-dependent. A quotient class may contain conflicting task labels ($\Delta_h > 0$) but have zero structural risk if the conflict occurs on a prior-measure-zero set.

**Proposition 11** (Structural Risk Characterization). *For 0-1 loss,*

$$R^*_Q = 0 \iff \text{almost every quotient class } q \text{ has } P(h(U) = y \mid Q = q) = 1 \text{ for some } y. \tag{18}$$

*Proof.* Let

$$e(q) := 1 - \max_{y \in \mathcal{H}} P(h(U) = y \mid Q = q). \tag{19}$$

Then $e(q) \geq 0$ and $R^*_Q = \mathbb{E}[e(Q)]$. A nonnegative random variable has zero expectation if and only if it is zero almost surely. Under 0–1 loss, $e(q) = 0$ if and only if the conditional task distribution given $Q = q$ is concentrated on one label, which proves the equivalence. □

Let $A_h = \{q \in \mathcal{Q}_O : P(h(U) = \cdot \mid Q = q) \text{ is non-degenerate}\}$. If $P_Q(A_h) > 0$, then $R^*_Q > 0$.

Indeed, $e(q) > 0$ on $A_h$. Since $\{e > 0\} = \bigcup_{n \geq 1} \{e \geq 1/n\}$, positive $P_Q$-measure of $A_h$ implies that $P_Q(e \geq 1/n) > 0$ for some $n$, and hence $\mathbb{E}[e(Q)] > 0$.

We use the total-variation convention

$$\mathrm{TV}(P_1, P_2) := \sup_{A \in \mathcal{A}_{\mathcal{Y}}} |P_1(A) - P_2(A)|, \tag{20}$$

which equals $\frac{1}{2} \|P_1 - P_2\|_1$ in the dominated case.

**Proposition 12** (Binary Two-State Impossibility). *Let $U \in \{u_1, u_2\}$ with equal prior, and $h(u_1) \neq h(u_2)$. Then the optimal observation-based Bayes risk is*

$$R^*_Y = \frac{1}{2}\left(1 - \mathrm{TV}(P_{Y|u_1}, P_{Y|u_2})\right). \tag{21}$$

*If $u_1 \sim_O u_2$, then $R^*_Y = 1/2$.*

*Proof.* Write $P_i = P_{Y|u_i}$. For a measurable decision region $A$ on which the rule predicts the second task label, its equal-prior risk is

$$\frac{1}{2}P_1(A) + \frac{1}{2}P_2(A^c) = \frac{1}{2}\left[1 - \left(P_2(A) - P_1(A)\right)\right]. \tag{22}$$

Because $P_1$ and $P_2$ have equal total mass, $\sup_A (P_2(A) - P_1(A)) = \mathrm{TV}(P_1, P_2)$. Minimizing over $A$ gives the stated risk. If $u_1 \sim_O u_2$, then $P_1 = P_2$, so the total variation is zero and $R^*_Y = 1/2$. □

### 3.5 Repetition Invariance

Let $Y^{(m)} = (Y_1, \ldots, Y_m)$ denote $m$ conditionally iid observations from the same fixed observation channel.

**Proposition 13** (Repetition Invariance of Structural Equivalence). *For any $m \geq 1$,*

$$u_1 \sim_O u_2 \iff P_{Y^{(m)}|u_1} = P_{Y^{(m)}|u_2}. \tag{23}$$

*Proof.* For conditionally iid repetitions of the same fixed channel, $P_{Y^{(m)}|u} = P_{Y|u}^{\otimes m}$. Equality of the one-sample laws therefore implies equality of their tensor-product laws. Conversely, equality of the $m$-sample laws implies equality of their first-coordinate marginals and hence of the one-sample laws. This argument requires only $m \geq 1$ and no domination assumption. □

**Remark 14.** *Under conditionally iid repetitions of the same fixed observation channel, exact structural equivalence is invariant to repetition. This does not preclude active sensing where the observation mechanism changes with actions.*

## 4 Local Statistical Geometry

### 4.1 Statistical Model Image

The observation mechanism $\mathcal{O}$ maps the latent space to a family of probability distributions:

$$\Theta_O = \{P_{Y|u} : u \in \mathcal{U}\} \subset \mathcal{M}(\mathcal{Y}). \tag{24}$$

Here $\mathcal{M}(\mathcal{Y})$ denotes the relevant set of probability measures on the observation space $\mathcal{Y}$.

The observation quotient is in bijection with the image of $\mathcal{O}$:

$$\mathcal{Q}_O \simeq \Theta_O, \tag{25}$$

under the identification $[u]_O \leftrightarrow P_{Y|u}$.

Explicitly, define $\iota : \mathcal{Q}_O \to \Theta_O$ by $\iota([u]_O) = P_{Y|u}$. It is well-defined by the definition of $\sim_O$, injective because equal image laws define the same quotient class, and surjective because $\Theta_O$ is the image family. This identification is set-theoretic; it does not by itself establish a measurable isomorphism, a smooth equivalence, or a diffeomorphism.

## 4.2 Fisher-Rao Geometry

For the local geometric analysis, restrict to a smooth parameter stratum $\mathcal{U}^{\mathrm{sm}} \subseteq \mathcal{U}$. Assume that the laws $P_{Y|u}$ are dominated by a common measure $\mu$, the densities $p(y|u)$ are differentiable in $u$, differentiation may be interchanged with integration where used, and directional scores are square-integrable. The latent parameterization may be redundant and Fisher-degenerate. We place nondegeneracy only on a smooth embedded statistical-image stratum $\Theta_{\mathcal{O}}^{\mathrm{reg}}$, on which the Fisher–Rao form is smooth and positive definite.

For tangent vectors $\xi, \zeta \in T_P\Theta_{\mathcal{O}}^{\mathrm{reg}}$ with score representations $s_\xi = d\xi/dP$ and $s_\zeta = d\zeta/dP$, the Fisher-Rao metric is

$$g_{\mathrm{FR},P}(\xi,\zeta) = \int_{\mathcal{Y}} s_\xi(y) s_\zeta(y)\, dP(y). \tag{26}$$

Equivalently, in parameter coordinates $u$,

$$G_{ij}(u) = \mathbb{E}_{Y\sim P_{Y|u}}\left[\partial_i \log p(Y|u)\, \partial_j \log p(Y|u)\right]. \tag{27}$$

## 4.3 Metric Descent to the Regular Quotient

**Theorem 15** (Fisher Geometry on the Regular Observation Quotient). *Assume:*

1. *The preceding local smooth dominated model and score assumptions hold on $\mathcal{U}^{\mathrm{sm}}$.*
2. *The regular stratum $\Theta_{\mathcal{O}}^{\mathrm{reg}} \subset \Theta_0$ is a smooth embedded statistical manifold on which $g_{\mathrm{FR}}$ is smooth and positive definite.*
3. *The map $\overline{\mathcal{O}} : \mathcal{Q}_{\mathcal{O}}^{\mathrm{reg}} \to \Theta_{\mathcal{O}}^{\mathrm{reg}}$, $[u] \mapsto P_{Y|u}$, is a diffeomorphism.*

*Then the pullback metric*

$$g_Q = \overline{\mathcal{O}}^* g_{\mathrm{FR}} \tag{28}$$

*defines a Riemannian metric on the regular observation quotient. Moreover, for $u \in \mathcal{U}^{\mathrm{sm}}$ whose observation law lies in the regular image,*

$$\ker G(u) = \ker d\mathcal{O}_u. \tag{29}$$

*Proof.* Because $\overline{\mathcal{O}}$ is a diffeomorphism, $d\overline{\mathcal{O}}_q$ is an isomorphism. Thus, for nonzero $v \in T_q\mathcal{Q}_{\mathcal{O}}^{\mathrm{reg}}$, $d\overline{\mathcal{O}}_q v \neq 0$ and

$$g_Q(v,v) = g_{\mathrm{FR}}(d\overline{\mathcal{O}}_q v, d\overline{\mathcal{O}}_q v) > 0. \tag{30}$$

Smoothness and symmetry also pull back, so $g_Q$ is Riemannian.

For $u \in \mathcal{U}^{\mathrm{sm}}$ and $v, w \in T_u\mathcal{U}^{\mathrm{sm}}$, the coordinate pullback identity is

$$(\mathcal{O}^* g_{\mathrm{FR}})_u(v,w) = \mathbb{E}_u\left[\partial_v \log p(Y|u)\, \partial_w \log p(Y|u)\right] \tag{31}$$

$$= v^T G(u) w. \tag{32}$$

Consequently,

$$v^T G(u) v = g_{\mathrm{FR}}(d\mathcal{O}_u v, d\mathcal{O}_u v). \tag{33}$$

Positive definiteness on the regular image gives $v^T G(u) v = 0$ if and only if $d\mathcal{O}_u v = 0$. Since $G(u)$ is positive semidefinite, $v^T G(u) v = 0$ if and only if $G(u)v = 0$. Hence $\ker G(u) = \ker d\mathcal{O}_u$. ∎

**Remark 16.** *Fisher-null directions are indistinguishable to first order. Global observational equivalence is distributional; Fisher degeneracy is a local differential property. These are distinct concepts.*

## 4.4 Target-Nuisance Decomposition

Assume the local parametric observation model

$$Y = f(x,\nu) + \epsilon, \qquad \epsilon \sim \mathcal{N}(0,\Sigma). \tag{34}$$

Here $\Sigma \succ 0$ is known and parameter-independent in the local model; these assumptions are required for the following mean-Jacobian Fisher formula. The Jacobians with respect to target and nuisance are

$$J_x = \frac{\partial f}{\partial x}(x_0,\nu_0), \qquad J_\nu = \frac{\partial f}{\partial \nu}(x_0,\nu_0). \tag{35}$$

The full Fisher information matrix for $(x,\nu)$ is

$$I = \begin{bmatrix} I_{xx} & I_{x\nu} \\ I_{\nu x} & I_{\nu\nu} \end{bmatrix} = \begin{bmatrix} J_x^T\Sigma^{-1}J_x & J_x^T\Sigma^{-1}J_\nu \\ J_\nu^T\Sigma^{-1}J_x & J_\nu^T\Sigma^{-1}J_\nu \end{bmatrix}. \tag{36}$$

The efficient Fisher information for $x$ obtained by removing nuisance-score directions has the generalized Schur form

$$I_{\mathrm{ef}} = I_{xx} - I_{x\nu} I_{\nu\nu}^{\dagger} I_{\nu x}. \tag{37}$$

Define the whitened target and nuisance Jacobians and the nuisance projector by

$$A := \Sigma^{-1/2} J_x, \qquad B := \Sigma^{-1/2} J_\nu, \qquad \boxed{\Pi_\nu := BB^{\dagger}}. \tag{38}$$

Then $I_{xx} = A^T A$, $I_{x\nu} = A^T B$, and $I_{\nu\nu} = B^T B$. The Moore–Penrose identity

$$B(B^T B)^{\dagger} B^T = BB^{\dagger} \tag{39}$$

shows that $\Pi_\nu$ is the Euclidean orthogonal projector onto $\mathrm{Col}(B)$. Therefore,

$$I_{\mathrm{ef}} = J_x^T \Sigma^{-1/2}(I - \Pi_\nu)\Sigma^{-1/2} J_x. \tag{40}$$

**Corollary 17** (Nuisance-Score-Projected Rank).

$$\ker(I_{\mathrm{ef}}) = \mathcal{N}_x(u_0), \tag{41}$$

*where*

$$\mathcal{N}_x(u_0) = \{\delta x \in T_{x_0}\mathcal{X} : \exists \delta\nu \in T_{\nu_0}\mathcal{N},\ J_x\delta x + J_\nu \delta\nu = 0\}. \tag{42}$$

*Proof.* For any target perturbation $\delta x$,

$$\delta x^T I_{\mathrm{ef}} \delta x = \|(I - \Pi_\nu) A\delta x\|_2^2. \tag{43}$$

Hence $\delta x \in \ker I_{\mathrm{ef}}$ if and only if $A\delta x \in \mathrm{Col}(B)$, equivalently if and only if there exists $\delta\nu$ such that $A\delta x + B\delta\nu = 0$, with the sign absorbed into $\delta\nu$. Since $\Sigma^{-1/2}$ is invertible, this is equivalent to $J_x\delta x + J_\nu\delta\nu = 0$. Thus $\ker I_{\mathrm{ef}} = \mathcal{N}_x(u_0)$. □

The kernel identity is pointwise and requires no constant-rank assumption. If the relevant Jacobian ranks are locally constant, the corresponding null spaces and projectors additionally admit the usual smooth local interpretation of constant dimension.

**Remark 18.** *Fisher-nullity characterizes first-order invisibility only. Higher-order terms may restore distinguishability in singular regimes; a systematic treatment lies beyond the scope of this work.*

## 5 Representation Accessibility

### 5.1 Representation Channels and Automatic Descent

Let $K(dz \mid y)$ be a state-independent encoder kernel; deterministic sample encoders are included as Dirac kernels. For $q = [u]_O$, set $P_{Y|q} := P_{Y|u}$ and define

$$R_K(q, B) := P_{Z|q}(B) = \int_{\mathcal{Y}} K(B \mid y)\, P_{Y|q}(dy). \quad (44)$$

The canonical encoder-induced object is therefore the quotient-indexed Markov kernel $R_K$; equivalently, it is the law family $q \mapsto P_{Z|q}$. It is well-defined because $P_{Y|q}$ is well-defined on quotient classes. For each measurable $B \subseteq Z$, the map

$$u \longmapsto \int_{\mathcal{Y}} K(B|y)\, O(u, dy) \quad (45)$$

is measurable and constant on observation-equivalence classes, so it descends measurably through Definition 4. Thus $q \mapsto R_K(q, B)$ is measurable.

**Proposition 19** (Automatic Quotient Descent of Post-Observation Representations)**.** *For a deterministic or stochastic encoder,*

$$u_1 \sim_O u_2 \implies P_{Z|u_1} = P_{Z|u_2}. \quad (46)$$

*Thus no observation-only post-processing can split an exact observation-equivalence class. This is the standard data-processing/garbling direction [3].*

*Proof.* For every measurable $B \subseteq Z$,

$$P_{Z|u}(B) = \int_{\mathcal{Y}} K(B \mid y)\, P_{Y|u}(dy). \quad (47)$$

If $u_1 \sim_O u_2$, the input measures are equal. Applying the same state-independent kernel $K$ therefore gives $P_{Z|u_1}(B) = P_{Z|u_2}(B)$ for every measurable $B$, hence equality of the output laws. □

The law family $R_K$ is the appropriate object for noisy observations, learned sample representations, Bayes risk, and statistical representation geometry. It does not automatically produce a deterministic point $\psi(q) \in Z$.

### 5.2 Fixed-Task Sufficiency

For a secondary deterministic geometric regime, separately assume or construct

$$\psi : Q_O^{\mathrm{eg}} \to Z. \quad (48)$$

It may be an oracle quotient coordinate, a canonical deterministic summary, a specified embedding of $P_{Z|q}$, or a controlled analytic representation. This idealized map is useful for geometric diagnostics and Experiment 3, but it is not the automatic output of an ordinary sample encoder.

Let a structurally identifiable task descend to

$$\bar{h} : Q_O^{\mathrm{eg}} \to H. \quad (49)$$

**Definition 20** (Exact Deterministic Task Sufficiency)**.** *The deterministic representation $\psi$ is sufficient for the fixed task, or has* task fidelity*, when*

$$\boxed{\psi(q_1) = \psi(q_2) \implies \bar{h}(q_1) = \bar{h}(q_2).} \quad (50)$$

*Equivalently,* $\sim_\psi \subseteq \sim_{\bar{h}}$.

**Proposition 21** (Standard Task-Decoder Factorization)**.** *Exact deterministic task sufficiency holds if and only if there is a well-defined decoder*

$$r_h : \psi(Q_O^{\mathrm{eg}}) \to H \quad (51)$$

*such that* $\bar{h} = r_h \circ \psi$.

*Proof.* The forward direction defines $r_h(\psi(q)) = \bar{h}(q)$; task sufficiency makes this independent of the preimage. The reverse direction follows immediately from the factorization. □

This is the standard constancy-on-fibers criterion, aligned with task-specific sufficiency and relevant-information traditions [17, 18, 20]. Injectivity of $\psi$ remains sufficient for every task, but is stronger than needed for one task:

$$\boxed{\begin{array}{c}\text{full quotient preservation}\\ \text{is stronger than fixed-task sufficiency}\end{array}} \quad (52)$$

### 5.3 Task-Decoder Geometric Accessibility

Fix metrics $d_Z$ and $d_H$. Under task sufficiency, define the task-decoder Lipschitz modulus

$$\boxed{\begin{aligned} L_{\mathrm{acc}}(h, \psi; d_Z, d_H) &:= \mathrm{Lip}(r_h) \\ &= \sup_{\psi(q_1) \neq \psi(q_2)} \frac{d_H(\bar{h}(q_1), \bar{h}(q_2))}{d_Z(\psi(q_1), \psi(q_2))} \end{aligned}} \quad (53)$$

Thus $L_{\mathrm{acc}} \in [0, \infty]$ is an extended Lipschitz modulus: task sufficiency makes the decoder well-defined but does not guarantee that this modulus is finite. We suppress $d_Z, d_H$ only when they are fixed. If task sufficiency fails, there is no induced deterministic task decoder; this exact failure must be reported separately from an infinite modulus caused by poor conditioning.

**Remark 22** (Metric and Scale Dependence)**.** *$L_{\mathrm{acc}}$ is metric-dependent and scale-conditional, not an intrinsic representation invariant. In Euclidean representation coordinates, for $c \neq 0$,*

$$L_{\mathrm{acc}}(h, c\psi) = |c|^{-1} L_{\mathrm{acc}}(h, \psi). \quad (54)$$

*Representation comparisons therefore require common predeclared metrics, task units, and a common scale convention. The modulus measures stability of the exact induced decoder; it does not by itself determine network size, optimization difficulty, or sample complexity.*

Let $d_Q$ be a declared metric on $Q_O^{\mathrm{eg}}$. When $d_Q$ is induced by the Fisher–Rao metric $g_Q$, distances are taken within a connected regular component using its geodesic distance; this does not define finite distances across disconnected components or include singular strata.

All suprema involving $d_Q$ below are understood over the declared metric domain. When $d_Q$ is Fisher-geodesic distance, this means a single connected regular component.

Define the task-agnostic quotient distortion

$$D_{\mathrm{geo}}(\psi) = \sup_{q_1 \neq q_2} \left| \frac{d_Z(\psi(q_1), \psi(q_2))}{d_Q(q_1, q_2)} - 1 \right|. \tag{55}$$

**Theorem 23** (Task-Agnostic Distortion Bound). *Let $\bar{h}$ be $L_{\bar{h}}$-Lipschitz with respect to $(d_Q, d_H)$. If*

$$D_{\mathrm{geo}}(\psi) \leq \epsilon < 1, \tag{56}$$

*then $\psi$ is injective,*

$$d_Z(\psi(q_1), \psi(q_2)) \geq (1 - \epsilon) d_Q(q_1, q_2), \tag{57}$$

*and its induced task decoder satisfies*

$$\boxed{L_{\mathrm{acc}}(h, \psi) \leq \frac{L_{\bar{h}}}{1 - \epsilon}.} \tag{58}$$

*Proof.* The distortion bound gives, for every distinct $q_1, q_2$ in the declared metric domain,

$$d_Z(\psi(q_1), \psi(q_2)) \geq (1 - \epsilon) d_Q(q_1, q_2). \tag{59}$$

Because $\epsilon < 1$ and $d_Q$ is a metric, the right-hand side is positive, so $\psi$ is injective. Moreover,

$$d_H(\bar{h}(q_1), \bar{h}(q_2)) \leq L_{\bar{h}} d_Q(q_1, q_2) \tag{60}$$

$$\leq \frac{L_{\bar{h}}}{1 - \epsilon} d_Z(\psi(q_1), \psi(q_2)). \tag{61}$$

Injectivity makes the induced decoder unambiguous. Taking the supremum over distinct represented points proves $L_{\mathrm{acc}}(h, \psi) \leq L_{\bar{h}}/(1 - \epsilon)$. □

This is an elementary inverse-Lipschitz stability bound, closely related to stable embedding theory [22]; it is not a new component theorem. It can be conservative because it protects quotient directions irrelevant to $\bar{h}$.

When $\bar{h} : Q_O^{\mathrm{reg}} \to H$ and $\psi : Q_O^{\mathrm{reg}} \to Z$ are smooth, and $(H, g_H)$ and $(Z, g_Z)$ are Riemannian, define the pullback forms

$$g_h = \bar{h}^* g_H, \qquad g_\psi = \psi^* g_Z \tag{62}$$

These forms measure first-order task relevance and deterministic representation geometry. A local task-conditioned requirement is

$$\ker g_\psi \subseteq \ker g_h, \tag{63}$$

or, more strongly and quantitatively, $g_h \preceq C^2 g_\psi$. Indeed, if $g_\psi(v, v) = 0$, then

$$0 \leq g_h(v, v) \leq C^2 g_\psi(v, v) = 0, \tag{64}$$

so the quantitative positive-semidefinite inequality implies the qualitative kernel inclusion. The converse kernel inclusion alone does not generally imply a uniform constant $C$. In particular, $g_h(v, v) = 0$ means first-order task nullity, not global task irrelevance.

## 5.4 Statistical and Information-Geometric Interface

The three geometric roles are distinct:

- $g_Q$: statistical distinguishability available from observation;
- $g_\psi$: geometry of an idealized deterministic representation;
- $g_h$: first-order task relevance.

There is no general ordering between $g_Q$ and an arbitrary $g_\psi$.

For the actual law-valued experiment $R_K$, assume dominated differentiability, interchange of differentiation with encoder integration, square-integrable scores, a state-independent encoder kernel, and regularity of the induced output statistical model. Then standard Fisher monotonicity gives

$$\boxed{g_Z^{\mathrm{FR}} \preceq g_Q} \tag{65}$$

For a tangent direction $v$, the output score satisfies

$$s_Z(Z) = \mathsf{E}[s_Y(Y) \mid Z]. \tag{66}$$

Conditional Jensen therefore yields

$$\mathsf{E}[s_Z(Z)^2] = \mathsf{E}\left[\mathsf{E}[s_Y(Y) \mid Z]^2\right] \leq \mathsf{E}[s_Y(Y)^2]. \tag{67}$$

Thus $g_Z^{\mathrm{FR}}(v, v) \leq g_Q(v, v)$ directionwise and hence $g_Z^{\mathrm{FR}} \preceq g_Q$ [21]. The representation-law Fisher metric $g_Z^{\mathrm{FR}}$ must not be identified with the deterministic pullback $g_\psi$.

Accordingly, Representation Accessibility is an observation-quotient diagnostic synthesis: it separates exact fixed-task sufficiency, scale-controlled decoder stability, local observation/representation/task geometry, and finite-sample representation-channel risk. None of these standard component facts is claimed as a new theorem.

# 6 Layered Task Risk Decomposition

## 6.1 Information Levels

Throughout this section, assume that the relevant conditional laws exist, the task and loss are measurable, the displayed risks are finite, admissible decision rules are measurable, and measurable Bayes rules or measurable $\eta$-optimal rules for every $\eta > 0$ exist at the information levels considered. The learned predictor is an admissible measurable rule based on $Z^{(m)}$.

Let $Y^{(m)} = (Y_1, \ldots, Y_m)$ denote $m$ conditionally iid observations from the same fixed channel. Let $K_m(dz \mid y^{(m)})$ be a state-independent representation kernel, and define the conditional representation law by

$$P_{Z^{(m)} \mid Y^{(m)} = y^{(m)}} = K_m(\cdot \mid y^{(m)}). \tag{68}$$

Deterministic representations are the special case $K_m(dz \mid y^{(m)}) = \delta_{\phi_m(y^{(m)})}(dz)$.

The relevant information levels are:

- $U$: full latent state.
- $Q = \pi_O(U)$: observation-law equivalence class (structural oracle).
- $Y^{(m)}$: finite observations from the fixed channel.
- $Z^{(m)}$: output of the state-independent representation kernel $K_m$ applied to $Y^{(m)}$.
- $\hat{h} = \hat{r}(Z^{(m)})$: learned measurable decision rule based on $Z^{(m)}$.

Here $Z^{(m)}$ is the actual random output of the representation channel. The deterministic quotient map $\psi$ used for geometric diagnostics in Section 5 is not automatically the same object; connecting it to $Z^{(m)}$ requires an explicit construction.

For $q = [u]_O$, define the quotient-indexed observation kernel by

$$\overline{O}(q, A) := O(u, A). \tag{69}$$

This is well-defined because $O(u, A)$ is constant on observation-equivalence classes. For each measurable $A$, the map $u \mapsto O(u, A)$ is measurable and fiber-constant, so Definition 4 gives measurable descent in $q$. Thus

$$O(u, A) = \overline{O}(\pi_O(u), A), \tag{70}$$

and the conditional observation law depends on $U$ only through $Q$. The induced iid product kernel therefore gives $U \to Q \to Y^{(m)}$. Applying the state-independent kernel $K_m$ gives the full chain $U \to Q \to Y^{(m)} \to Z^{(m)}$; deterministic $\phi_m$ is included as the Dirac-kernel special case.

The optimal risks at each level are:

$$R^*_U = \inf_{r: U \to H} \mathsf{E}[\ell(r(U), h(U))], \tag{71}$$

$$R^*_Q = \inf_{\bar{r}: Q_O \to H} \mathsf{E}[\ell(\bar{r}(Q), h(U))], \tag{72}$$

$$R^*_{Y^{(m)}} = \inf_{g: Y^m \to H} \mathsf{E}[\ell(g(Y^{(m)}), h(U))], \tag{73}$$

$$R^*_{Z^{(m)}} = \inf_{r: Z \to H} \mathsf{E}[\ell(r(Z^{(m)}), h(U))]. \tag{74}$$

**Proposition 24** (Monotonicity Across Information Levels). *Under the preceding decision-theoretic assumptions,*

$$R^*_U \le R^*_Q \le R^*_{Y^{(m)}} \le R^*_{Z^{(m)}} \le R(\hat{h}). \tag{75}$$

*Consequently, the adjacent gaps defined below are nonnegative and their sum is $R(\hat{h}) - R^*_U$.*

*Proof.* Because $Q = \pi_O(U)$ is a measurable function of $U$, every $Q$-based rule induces a $U$-based rule by composition, so $R^*_U \le R^*_Q$.

For an action $a$, define

$$L(q, a) := \mathsf{E}[\ell(a, h(U)) \mid Q = q]. \tag{76}$$

The Markov chain $U \to Q \to Y^{(m)}$ makes $Y^{(m)}$ conditionally independent of $h(U)$ given $Q$. Hence, for every $Y^{(m)}$-based rule $g$,

$$\mathsf{E}[\ell(g(Y^{(m)}), h(U)) \mid Q = q] \tag{77}$$

$$= \int L(q, g(y))\, P_{Y^{(m)}|q}(dy) \ge \inf_a L(q, a). \tag{78}$$

The measurable Bayes-rule assumption, or the $\eta$-optimal assumption followed by $\eta \downarrow 0$, then yields $R^*_Q \le R^*_{Y^{(m)}}$.

For an action $a$, define

$$L_Y(y, a) := \mathsf{E}[\ell(a, h(U)) \mid Y^{(m)} = y]. \tag{79}$$

Because $U \to Y^{(m)} \to Z^{(m)}$ under state-independent post-processing, $Z^{(m)}$ is conditionally independent of $h(U)$ given $Y^{(m)}$. Hence, for every measurable $Z^{(m)}$-based rule $r$,

$$\mathsf{E}[\ell(r(Z^{(m)}), h(U)) \mid Y^{(m)} = y] \tag{80}$$

$$= \int L_Y(y, r(z))\, K_m(dz \mid y) \ge \inf_a L_Y(y, a). \tag{81}$$

The measurable Bayes-rule assumption, or the $\eta$-optimal assumption followed by $\eta \downarrow 0$, then yields $R^*_{Y^{(m)}} \le R^*_{Z^{(m)}}$. Finally, $\hat{h}$ is one admissible $Z^{(m)}$-based predictor, giving $R^*_{Z^{(m)}} \le R(\hat{h})$.

Thus every adjacent difference is nonnegative, and direct cancellation gives

$$(R^*_Q - R^*_U) + (R^*_{Y^{(m)}} - R^*_Q) + (R^*_{Z^{(m)}} - R^*_{Y^{(m)}}) \tag{82}$$

$$+ (R(\hat{h}) - R^*_{Z^{(m)}}) = R(\hat{h}) - R^*_U. \tag{83}$$

□

## 6.2 Layered Risk Decomposition

For organizational purposes, we decompose the total risk gap as:

$$\boxed{R(\hat{h}) - R^*_U = \epsilon_{\text{struct}} + \epsilon_{\text{finite}}(m) + \epsilon_{\text{rep}} + \epsilon_{\text{learn}}} \tag{84}$$

where

$$\epsilon_{\text{struct}} = R^*_Q - R^*_U, \tag{85}$$

$$\epsilon_{\text{finite}}(m) = R^*_{Y^{(m)}} - R^*_Q, \tag{86}$$

$$\epsilon_{\text{rep}} = R^*_{Z^{(m)}} - R^*_{Y^{(m)}}, \tag{87}$$

$$\epsilon_{\text{learn}} = R(\hat{h}) - R^*_{Z^{(m)}} \ge 0. \tag{88}$$

Proposition 24 gives the nonnegativity of all four gaps and the displayed telescoping identity.

## 6.3 Interpretation

- $\epsilon_{\text{struct}}$: Irreducible loss due to exact observational collapse relative to the declared fixed observation channel. It cannot be reduced by additional conditionally iid repetitions of that same channel or by state-independent post-processing of it; changing the observation mechanism may change the quotient and hence the structural term.
- $\epsilon_{\text{finite}}(m)$: Loss due to finite-sample statistical overlap between distinguishable distributions. Under suitable separation conditions (e.g., finite task classes with distinct observation laws), $\epsilon_{\text{finite}}(m) \to 0$ as $m \to \infty$.
- $\epsilon_{\text{rep}}$: Task-, prior-, loss-, and sample-dependent loss due to the actual representation channel.

- $\epsilon_{\text{learn}}$: Gap between the selected predictor and the optimal rule from $Z^{(m)}$, including estimation, approximation, optimization, regularization, and misspecification effects as applicable.

**Corollary 25** (Structural Risk Floor). *If $A_h = \{q : P(h(U) = \cdot \mid Q = q)$ is non-degenerate$\}$ has positive $P_Q$-measure, then $\epsilon_{\text{struct}} > 0$ under 0-1 loss.*

*Proof.* By Proposition 11, $P_Q(A_h) > 0$ implies $R_Q^* > 0$. Since $h$ is measurable and deterministic and the rule $r = h$ is admissible, $R_U^* = 0$ under 0–1 loss. Therefore $\epsilon_{\text{struct}} = R_Q^* - R_U^* = R_Q^* > 0$. ⊓

# 7 Experimental Evaluation

## 7.1 Evaluation Design

The experiments are organized to evaluate the framework layer by layer rather than to optimize a single aggregate benchmark. Controlled synthetic studies isolate structural ambiguity, nuisance-efficient local information, task-relative representation accessibility, and layered risk attribution. We then examine the same diagnostic principles in a physics-based near-field sensing model and on frozen learned representations from NYUv2.

Across the controlled studies, the observation model, representation family, or task is varied only when required by the stated hypothesis. In particular, additional conditionally iid repetitions of one fixed channel cannot remove an exact quotient collision, whereas changing the observation mechanism may refine the quotient. The evidence therefore proceeds from structural availability through local statistical strength and representation accessibility to end-to-end risk, before testing the framework in physical and real-data settings.

## 7.2 Controlled Synthetic Validation

### 7.2.1 Structural Identifiability under Fixed and Changed Channels

We first test whether repetition separates finite statistical uncertainty from exact structural information loss. Let

$$U = (A, B), \qquad A, B \overset{\text{iid}}{\sim} \text{Bernoulli}(1/2), \tag{89}$$

and, conditionally on $A$, let repeated observations be iid with

$$\begin{aligned} Y_i \mid A = 0 &\sim \text{Bernoulli}(0.2), \\ Y_i \mid A = 1 &\sim \text{Bernoulli}(0.8). \end{aligned} \tag{90}$$

The channel is independent of $B$ given $A$, so exact enumeration gives the observation quotient $Q = A$. We compare

$$\begin{aligned} h_A(A, B) = A, \qquad & h_B(A, B) = B, \\ h_{AB}(A, B) = (A, B). & \end{aligned} \tag{91}$$

Here $h_A$ is quotient-compatible, $h_B$ is structurally incompatible, and $h_{AB}$ combines observable and hidden task content.

For $m \in \{1, 2, 4, 8, 16, 32, 64\}$ repetitions, exact Binomial likelihood calculations give

$$\begin{aligned} R^*_{Y^{(m)}}(h_A) &= e_m, \\ R^*_{Y^{(m)}}(h_B) &= \frac{1}{2}, \\ R^*_{Y^{(m)}}(h_{AB}) &= \frac{1}{2} + \frac{1}{2} e_m, \end{aligned} \tag{92}$$

where $e_m$ decreases from 0.2 at $m = 1$ to $5.07 \times 10^{-8}$ at $m = 64$. Repetition therefore drives the observable-task risk toward zero, leaves the hidden-task risk exactly at 1/2, and drives the mixed task to the same hidden-component floor. Their $(\epsilon_{\text{struct}}, \epsilon_{\text{finite}})$ pairs are $(0, e_m)$, $(1/2, 0)$, and $(1/2, e_m/2)$.

Using only the sufficient count $S_m = \sum_i Y_i$, a conditional-frequency learner trained on 50,000 examples over ten seeds tracks the analytic Bayes curves across the full repetition grid on independent test data (Fig. 1). This agreement recovers the distinct finite and structural limits for all three tasks: learning reduces observable uncertainty but does not cross the fixed-channel structural floor.

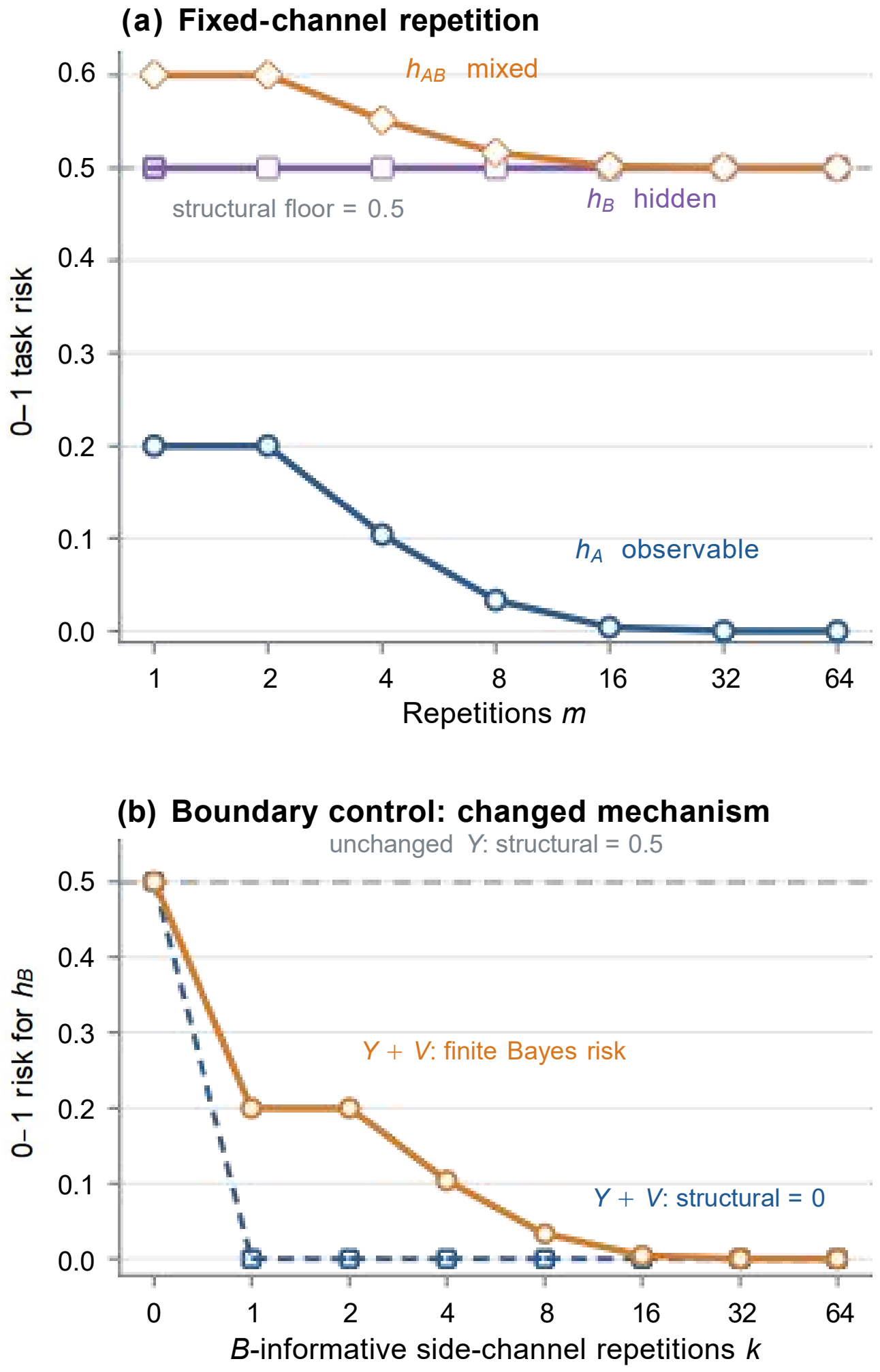

Figure 1: Structural identifiability under fixed and changed mechanisms. (a) Analytic Bayes risks and ten-seed learner intervals: repetition removes finite uncertainty for $h_A$, while $h_B$ and the hidden component of $h_{AB}$ remain at 0.5. (b) A $B$-informative side channel removes the collision; finite-observation error remains at finite $k$.

**Observation-only capacity control.** At $m = 16$, increasing interval-class capacity reduces mean $h_A$ risk from 0.4999 to 0.00425 and $h_{AB}$ risk from 0.7500 to 0.5016, while the $h_B$ means remain in $[0.4993, 0.4997]$ across ten seeds. The observable-task endpoints approach their analytic limits, whereas the hidden-task risk remains at its predicted floor. Capacity therefore removes observable approximation error in this family but cannot cross the channel-induced structural floor.

**Changed-mechanism boundary control.** For $h_B$, repetitions of the unchanged $Y$ law leave the structural quotient risk exactly 0.5. Under the augmented law $Y + V$, with $V \mid B = 0 \sim \mathrm{Bernoulli}(0.2)$ and $V \mid B = 1 \sim \mathrm{Bernoulli}(0.8)$, the structural term is zero for every $k \geq 1$. Finite Bayes risk decreases from 0.2 at $k = 1, 2$ to 0.104 at $k = 4$, 0.03334 at $k = 8$, and $5.07 \times 10^{-8}$ at $k = 64$. Thus a task-relevant source can refine the quotient, whereas repeated use of $Y$ preserves exact equivalence; this is not an active-policy optimization experiment.

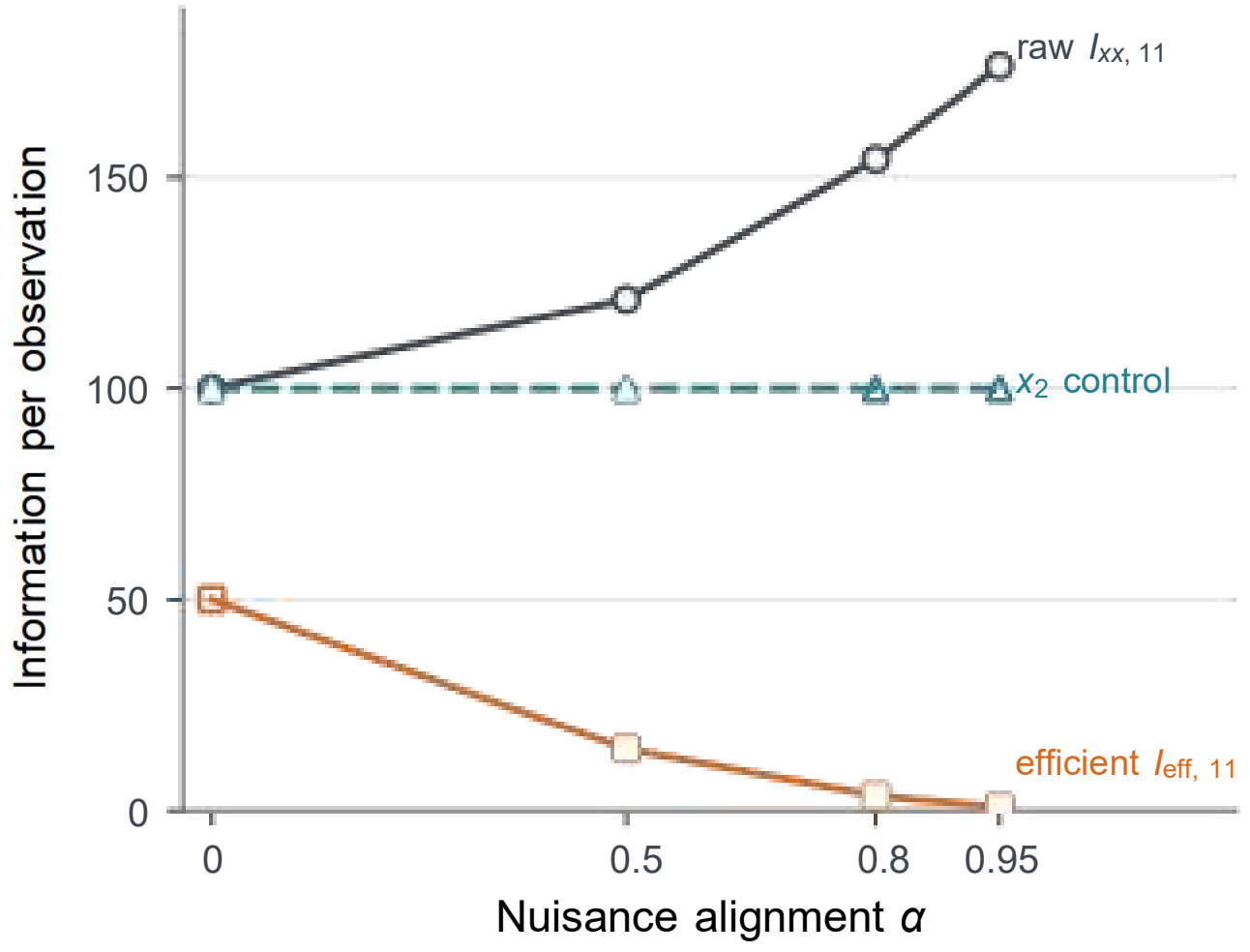

Figure 2: Analytic information under increasing target–nuisance alignment. The raw block $I_{xx,11}$ remains large, while nuisance projection collapses $I_{\mathrm{ef},11}$; the nuisance-orthogonal control $I_{\mathrm{ef},22}$ remains constant.

Changing only $P(B = 1) = \pi_B$ leaves the quotient unchanged but changes the hidden-task floor to $R^*_{\mathcal{Q}}(h_B) = \min(\pi_B, 1 - \pi_B)$, as the empirical risks track across the prior sweep. Structural compatibility is therefore channel–task dependent, while average structural risk also depends on the prior and loss.

### 7.2.2 Nuisance-Efficient Fisher Geometry

We next ask whether nuisance-efficient information predicts target-estimation behavior when target and nuisance score directions overlap. For known parameter-independent $\sigma = 0.1$, each observation follows

$$Y_i \mid (x_1, x_2, \nu) \sim \mathcal{N}\left(\begin{bmatrix} x_1 + \nu \\ x_2 \\ \nu + \alpha \sin x_1 \end{bmatrix}, \sigma^2 I_3\right). \tag{93}$$

The regular truth is $(x_1^\star, x_2^\star, \nu^\star) = (0.4, -0.3, 0.2)$, with $\alpha \in \{0, 0.5, 0.8, 0.95\}$. Analytic nuisance-score projection gives

$$I_{\mathrm{eff}} = \frac{1}{\sigma^2}\begin{bmatrix} (1 - \alpha\cos x_1)^2/2 & 0 \\ 0 & 1 \end{bmatrix}. \tag{94}$$

At this truth, $I_{\mathrm{ef},11}$ decreases from 50 to 0.7812 across the alignment grid, whereas $I_{\mathrm{ef},22} = 100$. Figure 2 shows that the raw target block remains large even as nuisance projection removes almost all efficient $x_1$ information.

Across ten seeds, 10,000 trials per condition, and $m \in \{8, 16, 32, 64, 128, 256, 512, 1024\}$, the regular study sampled the exact Gaussian distribution of the sample mean and used the model-specific maximum-likelihood reduction

$$\hat{x}_2 = \bar{Y}_2, \qquad \hat{x}_1 = g_\alpha^{-1}(\bar{Y}_1 - \bar{Y}_3), \qquad g_\alpha(x) = x - \alpha \sin x. \tag{95}$$

At $m = 1024$, the values of

$$Q_1(m) = m\,\mathrm{MSE}(\hat{x}_1) I_{\mathrm{ef},11} \tag{96}$$

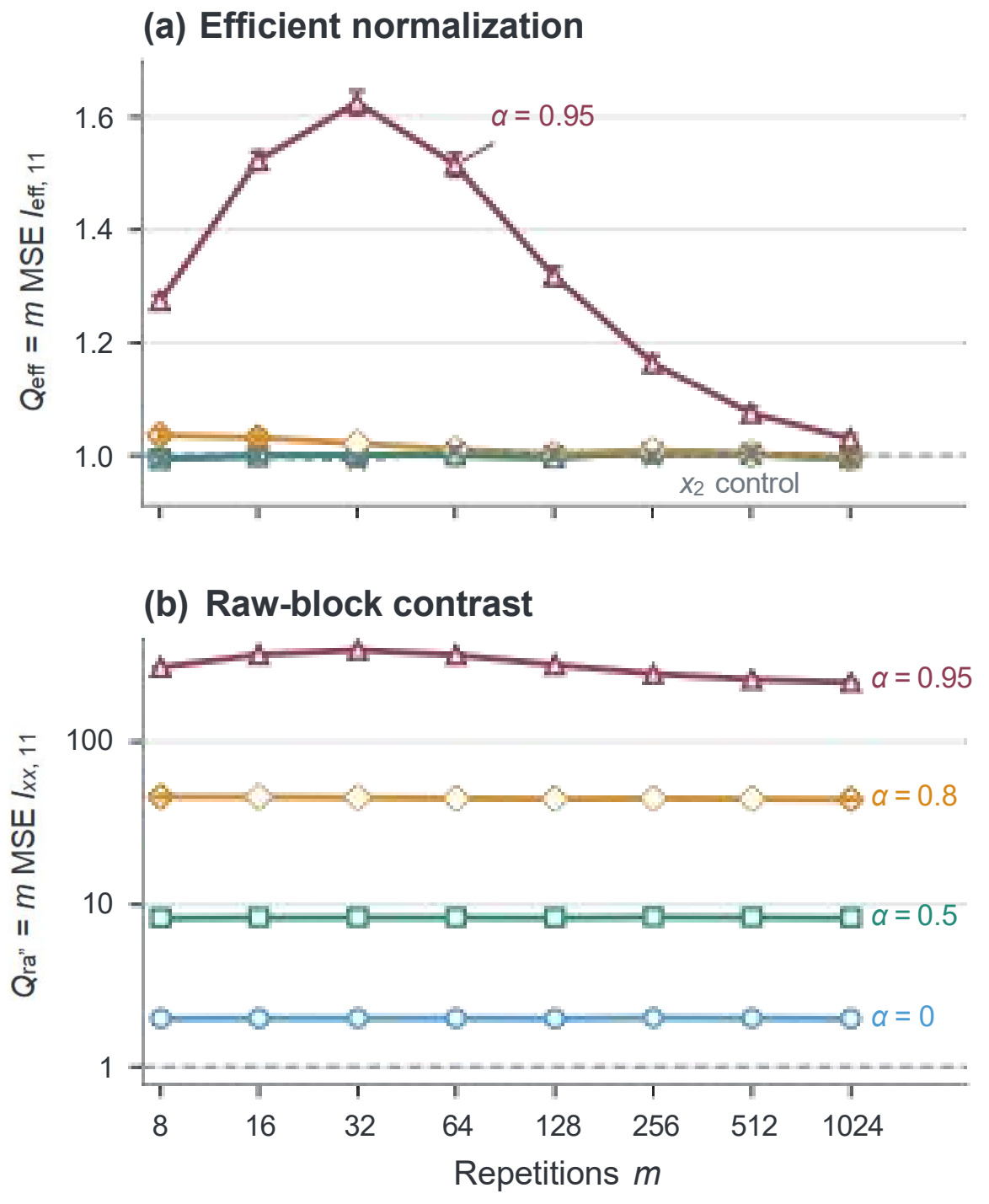

Figure 3: Fisher-normalized MLE error over ten seeds. Nuisance-efficient normalization approaches the asymptotic reference one across regular conditions, whereas raw-block normalization becomes increasingly miscalibrated with score alignment. Markers and capped bars denote means and normal-approximation 95% intervals across ten seed-level Monte Carlo batches.

range from 0.9947 to 1.029, close to the regular asymptotic reference $Q_1 = 1$. Nominal 95% efficient-information intervals achieve 0.9475–0.9509 empirical coverage at this largest sample size.

For the same estimator outputs, Fig. 3 shows raw-block normalization increasing from 1.989 to 232.6. The raw target block therefore becomes increasingly miscalibrated as the target score aligns with the nuisance direction. The comparison separates efficient local calibration from misleading raw-block scaling on identical estimator outputs.

A singular negative control sets $\alpha = 1$ and $x_1^\star = 0$. Here $I_{\mathrm{ef},11} = 0$, but

$$g_1(x) = x - \sin x = \frac{x^3}{6} + O(x^5), \tag{97}$$

so the state remains distinguishable beyond first order and regular root-$m$ asymptotics do not apply. The large-$m$ fitted MSE slope is $-0.3346$, consistent with the model-specific cubic prediction $-1/3$. Together, these results validate the nuisance-efficient local prediction in the declared model; Fisher rank remains a first-order local diagnostic rather than a global identifiability criterion.

### 7.2.3 Task-Relative Accessibility

This controlled comparison asks whether matched task-agnostic representation geometry implies matched fixed-task accessibility. We hold the quotient substrate, representation dimension, metric, and global scale fixed while changing only the downstream task. Let $Q = [0, 1]^2$ with the Euclidean metric and

$$\psi_A(q_1, q_2) = (1.5q_1, 0.5q_2),$$
$$\psi_B(q_1, q_2) = (0.5q_1, 1.5q_2). \tag{98}$$

The maps have the same dimension, singular values $\{1.5, 0.5\}$, Frobenius norm $\sqrt{5/2}$, determinant magnitude 3/4, and $D_{\mathrm{geo}} = 1/2$. Their RMS norms and covariance traces also match under the symmetric reference distribution; the Euclidean metric and scale are fixed across conditions.

For

$$h_1(q) = q_1, \qquad h_2(q) = q_2, \qquad h_3(q) = \frac{q_1 + q_2}{\sqrt{2}}, \tag{99}$$

the analytic task-decoder moduli are

$$L_{\mathrm{acc}}(h_1, \psi_A) = \frac{2}{3}, \qquad L_{\mathrm{acc}}(h_1, \psi_B) = 2, \tag{100}$$

$$L_{\mathrm{acc}}(h_2, \psi_A) = 2, \qquad L_{\mathrm{acc}}(h_2, \psi_B) = \frac{2}{3}, \tag{101}$$

$$L_{\mathrm{acc}}(h_3, \psi_A) = L_{\mathrm{acc}}(h_3, \psi_B) = \frac{\sqrt{20}}{3}. \tag{102}$$

The analytic prediction is therefore a reversal for $h_1$ versus $h_2$ and a tie for $h_3$, despite identical task-agnostic distortion.

We sampled $Q_1, Q_2 \overset{\mathrm{iid}}{\sim} \mathrm{Uniform}(0, 1)$ with 20,000 training and 100,000 test points per seed over ten seeds. The paired comparison used the same samples, Euclidean representation metric, scale convention, and decoder protocol for both representations, changing only $h$. We fit the declared restricted family

$$\mathcal{R}_\beta = \{r(z) = w^T z + b : \|w\|_2 \le \beta\} \tag{103}$$

by convex norm-constrained least squares. At the fixed budget $\beta = 1$, Fig. 4 and Table 1 show that $h_1$ favors $\psi_A$, $h_2$ favors $\psi_B$, and $h_3$ ties. The complete budget grid follows the analytic decoder-norm thresholds. Matched task-agnostic geometry therefore does not imply matched accessibility for a fixed task and decoder family.

Table 1: Mean held-out MSE at $\beta = 1$ over ten paired seeds; lower values indicate greater task accessibility.

| Task | $\psi_A$ | $\psi_B$ | Lower MSE |
|---|---|---|---|
| $h_1$ | $7.1 \times 10^{-30}$ | $2.086 \times 10^{-2}$ | $\psi_A$ |
| $h_2$ | $2.084 \times 10^{-2}$ | $3.6 \times 10^{-30}$ | $\psi_B$ |
| $h_3$ | $5.740 \times 10^{-3}$ | $5.745 \times 10^{-3}$ | tie |

**Controls and boundaries.** For representation perturbations $\delta \sim \mathcal{N}(0, \sigma^2 I)$, the exact linear decoder predicts

$$\mathrm{MSE} = \sigma^2 L_{\mathrm{acc}}^2. \tag{104}$$

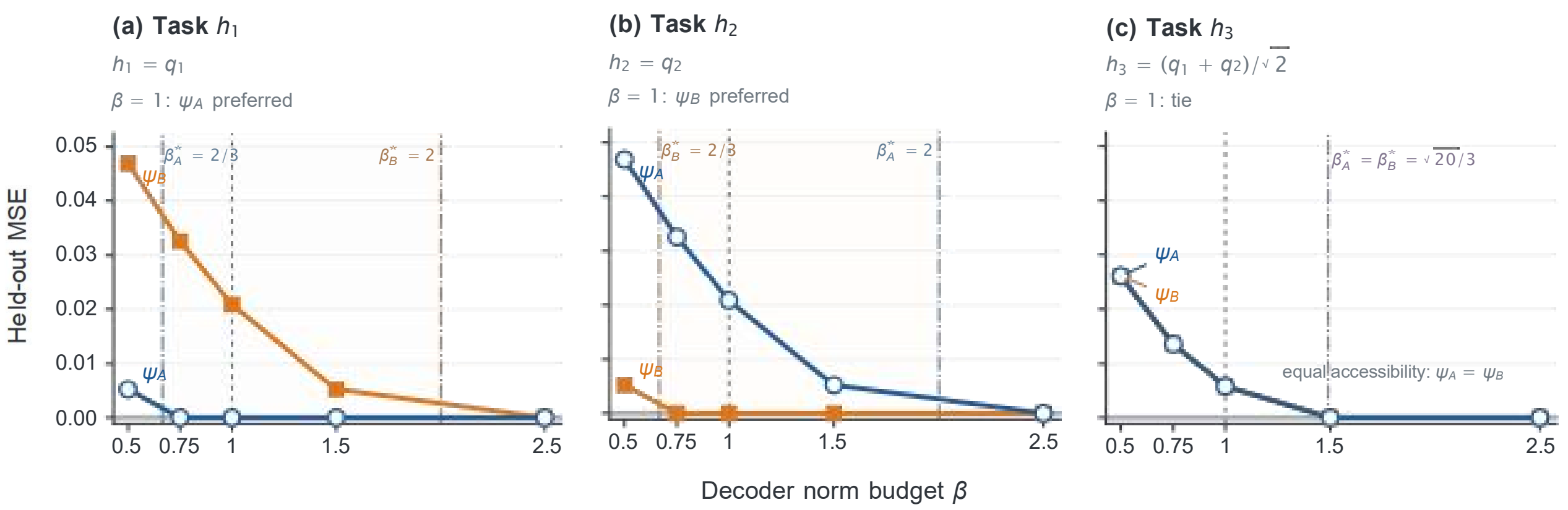

Figure 4: Task-relative accessibility under $R_\beta$ with representations and paired samples fixed. At $\beta = 1$, $h_1$ favors $\psi_A$, $h_2$ favors $\psi_B$, and $h_3$ ties. Points are ten-seed means, capped bars are normal-approximation 95% intervals, and dash-dotted lines mark analytic exact-decoder norms.

Table 2: Task-sufficiency control for $\psi_C(q) = q_1$. Ten-seed held-out MSE recovers $h_1$ exactly and matches the analytic information-loss references for $h_2$ and $h_3$.

| Task | Empirical MSE | Analytic reference | Interpretation |
|---|---|---|---|
| $h_1$ | $2.3 \times 10^{-32}$ | 0 | exact recovery |
| $h_2$ | 0.08337 | 1/12 | discarded $q_2$ |
| $h_3$ | 0.04168 | 1/24 | partial $q_2$ loss |

The perturbation and worst-direction controls follow this analytic scaling across the tested conditions.

First, $\psi_C(q_1, q_2) = q_1$ is exactly sufficient for $h_1$ without being injective on $Q$, establishing that task sufficiency does not require full quotient preservation. Table 2 gives numerical exact recovery for $h_1$ and the analytic information-loss floors for $h_2$ and $h_3$.

Second, quotient-irrelevant lifts $s_c^{(1)}(q) = (q_1, q_2, cq_1)$ and $s_c^{(2)}(q) = (q_1, q_2, cq_2)$ for $c \in \{0, 0.5, 1, 2, 4\}$ leave the quotient, tasks, representations, decoder moduli, and performance unchanged. Figure 5 shows $D^Q_{\text{geo}} = 0.5$ throughout, while the deliberately mismatched ambient control varies with the representative section and reverses the A/B ordering. Arbitrary representative ambient geometry therefore need not descend to an intrinsic quotient geometry; this control does not reject physical Euclidean geometry in general.

Third, the analytic boundary $\psi_k(q) = q^k$ on $[0, 1]$ is injective for $k > 1$ but satisfies

$$D_{\text{geo}}(\psi_k) = \max\{1, k - 1\} \geq 1, \tag{105}$$

and $\psi_k^{-1}(z) = z^{1/k}$ is not Lipschitz at zero. On $Q_\varepsilon = [\varepsilon, 1]$, the grid $k \in \{2, 3, 4\}$ and $\varepsilon \in \{10^{-1}, 5 \times 10^{-2}, 2 \times 10^{-2}, 10^{-2}, 5 \times 10^{-3}, 10^{-3}\}$ gives

$$L_{\text{acc}}(k, \varepsilon) = \frac{1}{k\varepsilon^{k-1}}. \tag{106}$$

over all 18 injective, task-sufficient conditions. The modulus ranges from 5 to $2.5 \times 10^8$ and agrees with the analytic maxima. Thus exact information preservation need not imply uniformly stable inversion; all tested conditions lie outside the positive $D_{\text{geo}} < 1$ regime.

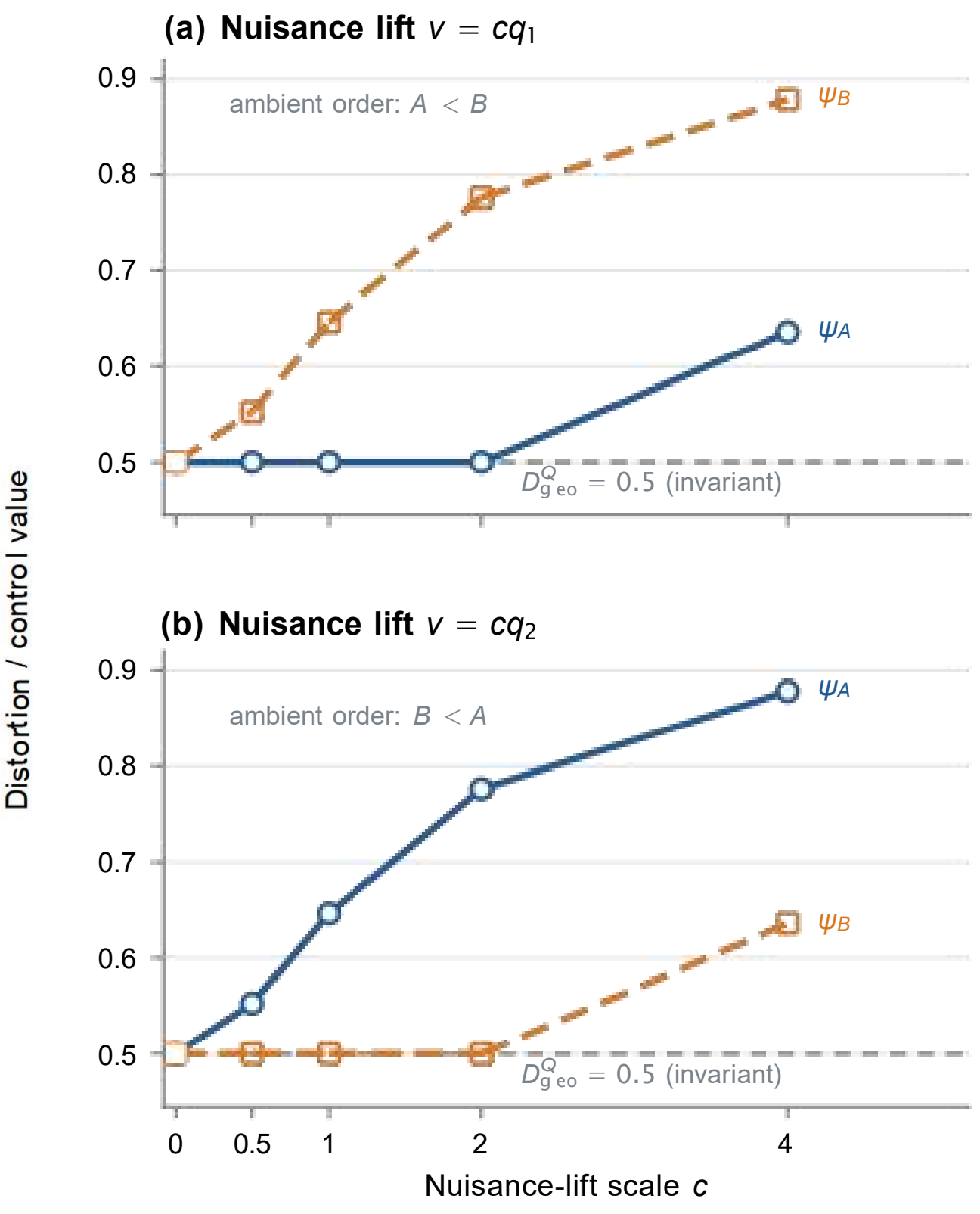

Figure 5: Quotient invariance under two quotient-irrelevant nuisance lifts. The declared distortion remains $D^Q_{\text{geo}} = 0.5$ for both representations, whereas the ambient-representative control varies with the section and reverses the A/B ordering.

#### 7.2.4 Layered Risk Attribution

We reuse the binary latent/channel construction above and set $h(U) = (A, B)$ under normalized Hamming loss

$$\ell((\hat{A}, \hat{B}), (A, B)) = \frac{1}{2}\mathbf{1}\{\hat{A} \neq A\} + \frac{1}{2}\mathbf{1}\{\hat{B} \neq B\}. \tag{107}$$

The observation channel still depends only on $A$, so $Q = A$. A state-independent representation channel applies

$$Z_i = Y_i \oplus E_i, \qquad E_i \sim \text{Bernoulli}(\gamma), \tag{108}$$

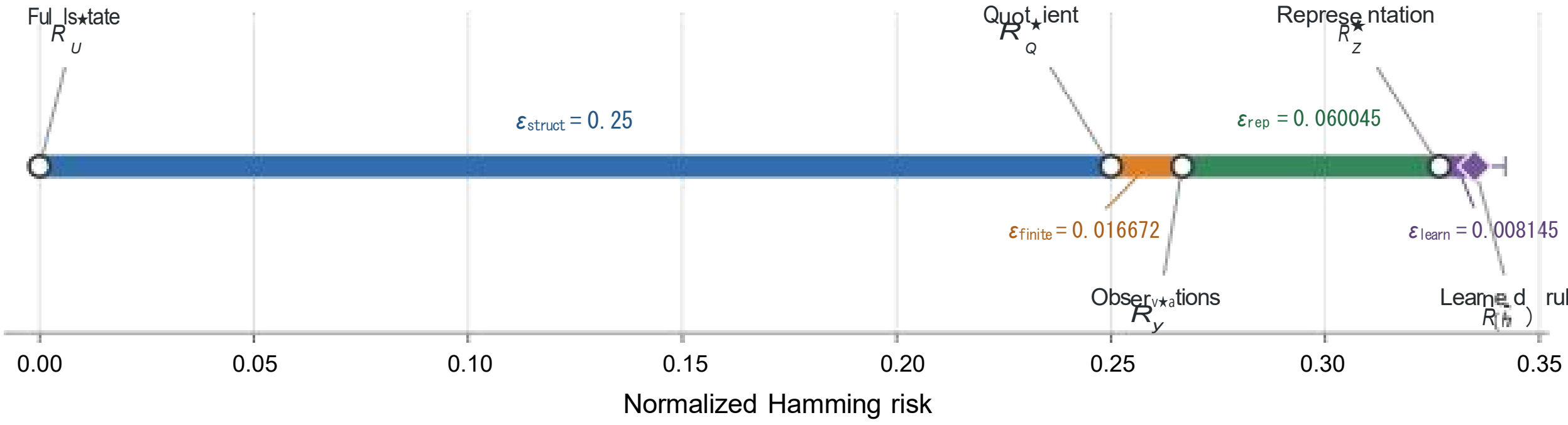

Figure 6: Cumulative layered-risk attribution. Five nodes mark full-state, quotient, finite-observation, representation-channel, and learned-rule risks; colored segments are the four adjacent gaps. The first four nodes are analytic oracle quantities, and the final diamond is the 20-seed learned-rule mean with a 95% interval.

independently, producing the information chain

$$U \longrightarrow Q \longrightarrow Y^{(m)} \longrightarrow Z^{(m)}. \tag{109}$$

For $m = 8$, $\gamma = 0.2$, $P(B = 1) = 0.5$, and $n_{\text{train}} = 50$, exact Binomial oracles and the learned conditional-frequency lookup give

$$\begin{aligned}(R^\star_U, R^\star_Q, R^\star_Y) &= (0,\ 0.25000,\ 0.26667),\\ (R^\star_Z, R(\hat{h})) &= (0.32672,\ 0.33486).\end{aligned} \tag{110}$$

The corresponding gaps are

$$\begin{aligned}&(\epsilon_{\text{struct}}, \epsilon_{\text{finite}}, \epsilon_{\text{rep}}, \epsilon_{\text{learn}})\\ &\quad = (0.25000,\ 0.01667,\ 0.06005,\ 0.00814).\end{aligned} \tag{111}$$

The four adjacent gaps isolate hidden $B$, finite uncertainty about $A$, bit-flip representation loss, and the finite-data lookup gap, respectively. Their sum is the exact telescoping decomposition in Section 6; Fig. 6 visualizes the cumulative attribution.

Four controlled sweeps isolate the terms by design. Varying only $\pi_B = P(B = 1)$ gives

$$\epsilon_{\text{struct}} = \frac{1}{2}\min\{\pi_B, 1 - \pi_B\}, \tag{112}$$

without changing the quotient; the finite and representation gaps remain unchanged to numerical precision. Varying $m$ with $\gamma = 0$ changes only $\epsilon_{\text{finite}}$, while varying $\gamma$ at $m = 8$ changes only $\epsilon_{\text{rep}}$ and gives zero representation gap at $\gamma = 0$.

The stochastic intervention fits $T_m = \sum_i Z_i \mapsto \hat{A}$ from finite training data and evaluates each rule under its exact population distribution. Across 20 seeds, the mean learning gap decreases from 0.02465 at $n_{\text{train}} = 20$ to zero by $n_{\text{train}} = 200$; Fig. 7 places all four interventions on one risk-gap scale.

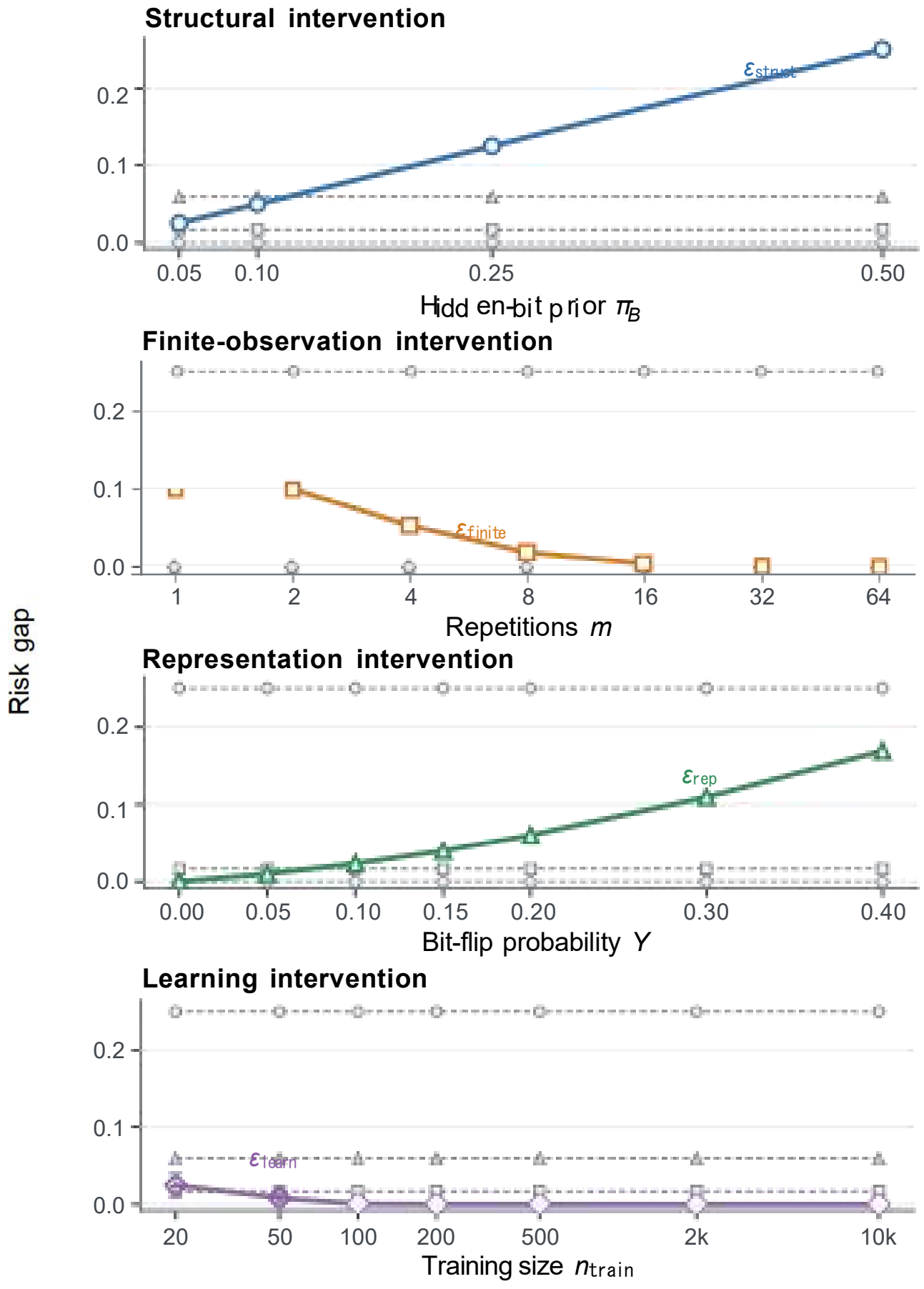

Figure 7: One-factor verification of layered risk. Each strip varies one control while displaying all four gaps on a common scale; the targeted gap is highlighted and non-target gaps are neutral references. The learning strip reports 20-seed means with 95% intervals where nonzero.

The construction demonstrates operational separability in this designed model:

$$\begin{aligned}\pi_B &\Rightarrow \epsilon_{\text{struct}}, & m &\Rightarrow \epsilon_{\text{finite}},\\ \gamma &\Rightarrow \epsilon_{\text{rep}}, & n_{\text{train}} &\Rightarrow \epsilon_{\text{learn}}.\end{aligned} \tag{113}$$

This one-factor manipulability is specific to the construction, not a general independence theorem.

### 7.3 Physics-Based Case Study: Near-Field Sensing

Can one sensing mechanism support different tasks with sharply different local statistical strength? We examine this question in a physics-based synthetic forward model, not a hardware experiment [33]. A 32-element half-wavelength ULA at $f_c = 30\text{GHz}$ observes

$$Y = \beta a(r,\theta) + W, \qquad W \sim \mathcal{CN}(0,\sigma^2 I), \tag{114}$$

where the unknown complex gain $\beta$ is nuisance and the constant-modulus spherical-wave steering law is

$$\begin{aligned} a_n(r,\theta) &= \exp(-\mathrm{j}kd_n(r,\theta)), \\ d_n(r,\theta) &= \sqrt{r^2 + x_n^2 - 2rx_n \sin\theta}. \end{aligned} \tag{115}$$

The physical aperture is $D = 0.1549\text{m}$ and the conventional Rayleigh reference distance is $r_R = 2D^2/\lambda = 4.802\text{m}$. We use $r_R$ only to normalize range, not as a sharp identifiability boundary.

For this standard proper-complex Gaussian specialization, nuisance-score projection gives

$$I_{\text{eff}} = \frac{2|\beta|^2}{\sigma^2}\,\mathrm{Re}\left[J_a^H P_a^{\perp} J_a\right], \qquad P_a^{\perp} = I - a(a^H a)^{-1}a^H. \tag{116}$$

The numerical Fisher sweep treats range and angle as two tasks under the same array and unknown-gain model [34]. Table 3 and Fig. 8 show rapid degradation of range support with distance, while angular support remains approximately stable in this configuration.

Table 3: Near-field task information and one-snapshot CRLB standard deviations. Each scalar task treats the other target coordinate as unknown; range and angle retain their physical units.

| $r/r_R$ | $I_r^{\text{task}}$ | $\text{SD}_r$ (m) | $I_\theta^{\text{task}}$ | $\text{SD}_\theta$ (deg) |
|---|---|---|---|---|
| 0.25 | 85.54 | 0.1081 | $4.728 \times 10^5$ | 0.08333 |
| 1 | 0.3352 | 1.727 | $4.753 \times 10^5$ | 0.08311 |
| 4 | $1.309 \times 10^{-3}$ | 27.63 | $4.755 \times 10^5$ | 0.08309 |

Over the far-field tail, the fitted range-information log–log slope is $-4.0002 \pm 0.0002$, consistent with the model-specific expansion $I_r^{\text{task}} \propto r^{-4}$ for this centered narrowband ULA.

**Profile-MLE/CRLB study.** The profile MLE globally searches $(r,\theta)$ by coarse-to-fine minimization of

$$\left\|\bar{Y} - \hat{\beta}(r,\theta)a(r,\theta)\right\|_2^2, \qquad \hat{\beta}(r,\theta) = \frac{a(r,\theta)^H \bar{Y}}{a(r,\theta)^H a(r,\theta)}. \tag{117}$$

The study uses ten seeds, 10,000 trials per range–snapshot condition, and $m \in \{1, 4, 16, 64, 256\}$. At $m = 256$, the global normalized MSEs

$$Q_r = m\,\mathrm{MSE}(\hat{r})I_r^{\text{task}}, \qquad Q_\theta = m\,\mathrm{MSE}(\hat{\theta})I_\theta^{\text{task}} \tag{118}$$

are (0.988, 0.993) at $0.25r_R$ and (0.989, 0.981) at $r_R$, with 100% correct-basin rates. These regular, well-resolved conditions approach the nuisance-efficient CRLB.

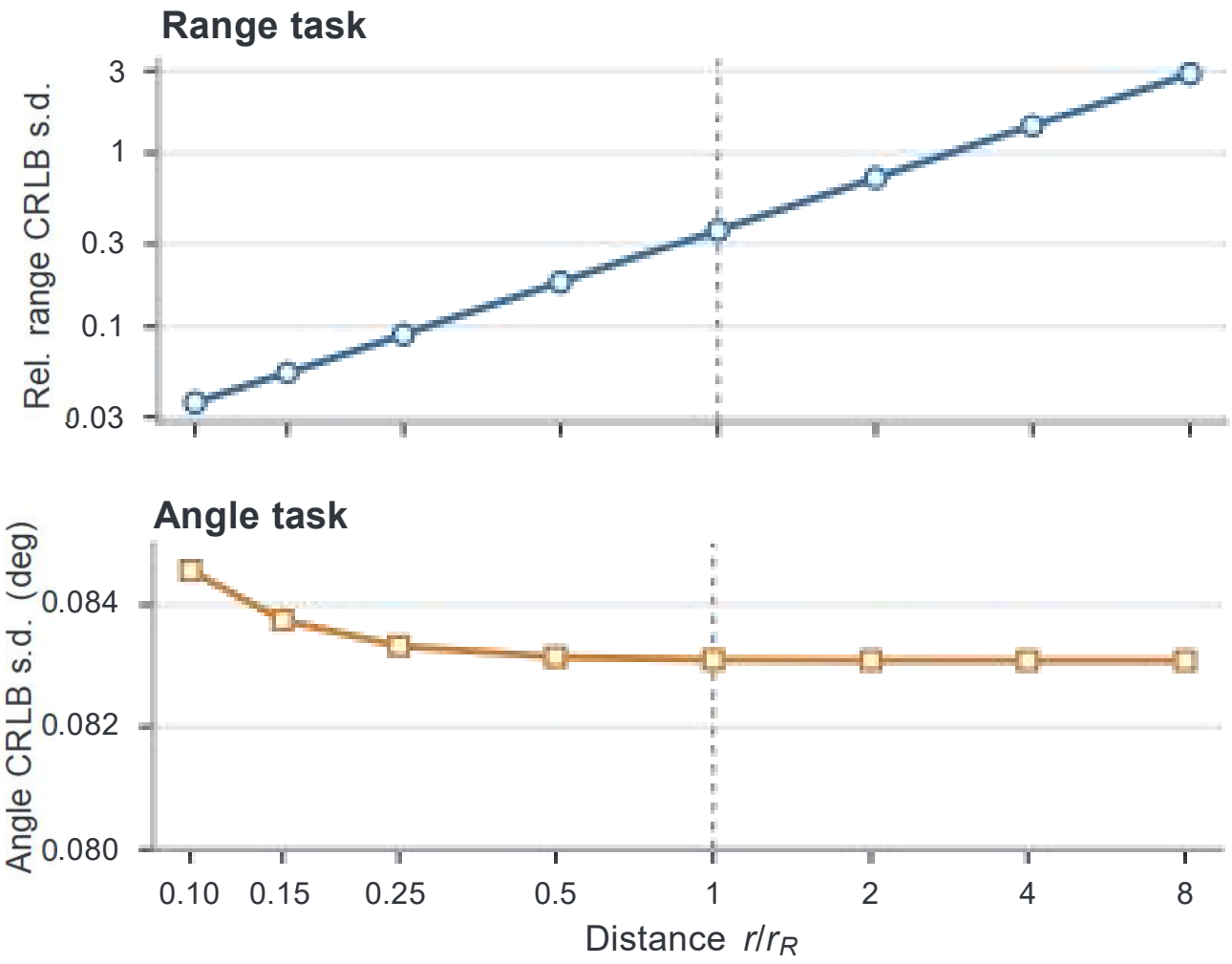

Figure 8: Task-specific CRLB standard deviations versus normalized distance under the spherical-wave model. Range uncertainty grows rapidly, while angular uncertainty remains nearly stable. The dashed line at $r/r_R = 1$ is the conventional Rayleigh normalization reference, not an identifiability boundary.

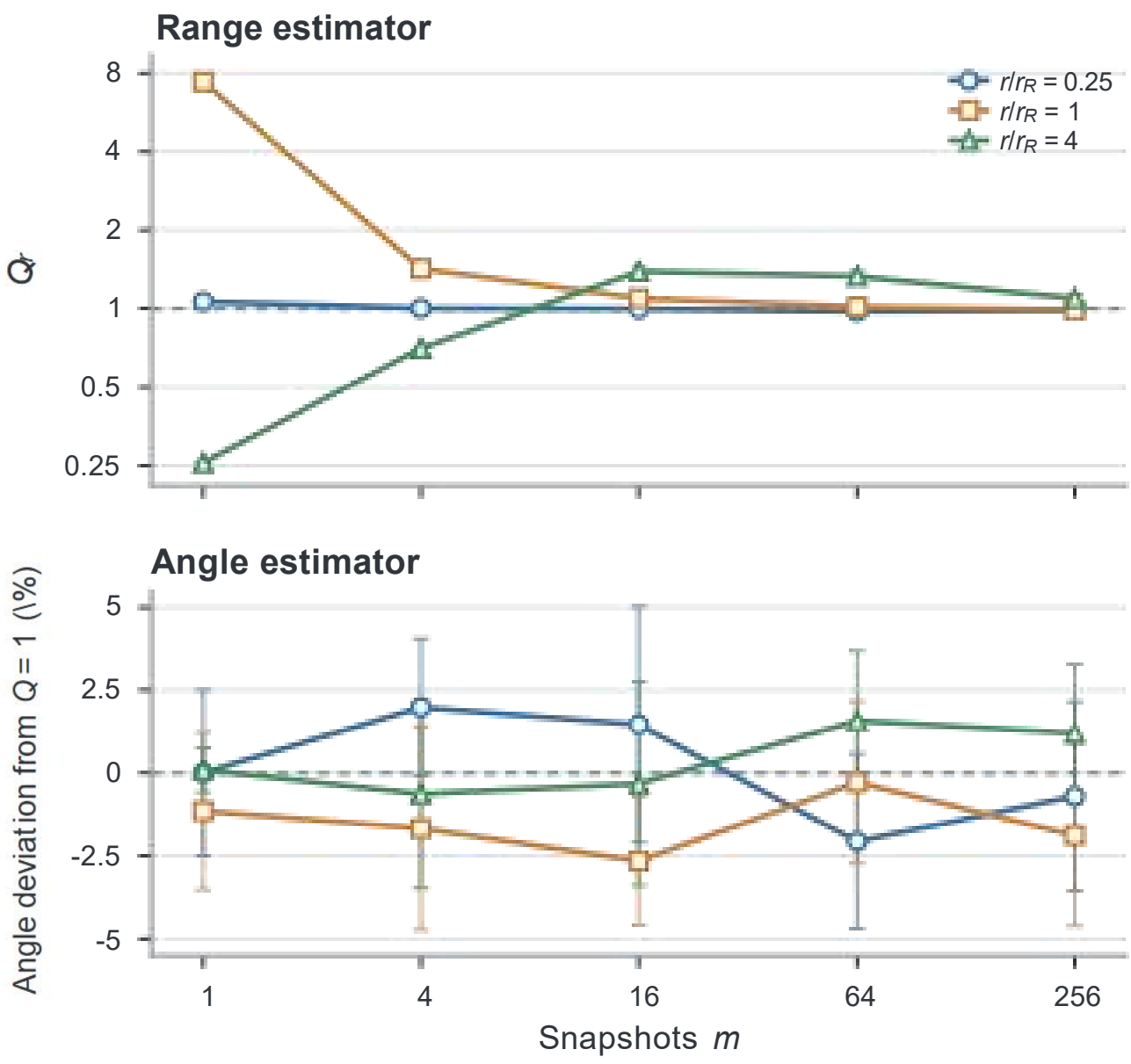

Figure 9: Global profile-MLE Fisher-normalized error. The upper strip shows $Q_r$ for three ranges; the lower strip shows $100(Q_\theta - 1)\%$. Curves are ten-seed means with 95% intervals. The $0.25r_R$ and $r_R$ cases are regular and well resolved, whereas $4r_R$ exposes weak-information and global-estimation effects.

At $4r_R$, $(Q_r, Q_\theta) = (1.094, 1.012)$, but the correct-basin rate falls to 73.35%, with 1.31% gross aliases and 5.95% optimizer diagnostic flags. Figure 9 therefore treats this case as a weak-information/global-estimation diagnostic: local Fisher geometry does not characterize global likelihood ambiguity.

**Plane-wave negative control.** In the ideal narrowband constant-modulus plane-wave model, after common phase is absorbed into unknown $\beta$, the steering law has no range dependence. Under this control, $\|\partial_r a^{FF}\| = 0$, $I_r^{task} = 0$, and the profiled likelihood is constant in range, while angle information remains approximately $4.755 \times 10^5$. This establishes only that range is absent from this declared plane-wave observation model; it does not imply that all far-field sensing systems lack range information.

## 7.4 Real-World Evaluation on NYUv2

Do task-relative representation preferences persist for real RGB observations and frozen learned encoders, first under a common controlled affine probe and then under a second pointwise decoder family? We use the official NYUv2 train/validation/test scene split with no scene leakage. Final post-LayerNorm patch-token features from architecture-matched DeiT-B/16, DINO ViT-B/16, and MAE ViT-B/16 encoders are frozen and processed identically. NYU40 semantic segmentation and metric-depth regression use the same task-appropriate $1 \times 1$ affine head family, with mIoU and AbsRel as their respective metrics and ten probe seeds per encoder–task pair.

**Full-data task interaction.** Mean mIoU is 0.3400, 0.3187, and 0.3110 for DeiT, DINO, and MAE; mean depth AbsRel is 0.2300, 0.2289, and 0.1905 (lower is better). The decisive DeiT–MAE contrast favors DeiT for semantics, +0.0434 with 95% interval [0.0244, 0.0603], but favors MAE for depth, −0.1877 with interval [−0.2184, −0.1579]. Its cross-task interaction is 0.2311 with familywise interval [0.1887, 0.2724]. Figure 10 shows the task-specific affine contrasts, while Table 4 reports exact cross-task interactions under the affine and MLP-128 probes.

**Probe-family robustness.** To test decoder-family dependence, we added a pointwise MLP-128 head, $768 \to 128 \to C$ with GELU, with frozen features, an unchanged spatial interface, no spatial mixing, and no encoder fine-tuning. It received the same taskwise validation-only learning-rate/weight-decay selection opportunity as the affine family, using one setting per task across encoders and ten confirmatory seeds. For DeiT–MAE, the MLP contrast remains positive for semantics, +0.0375 with paired 95% interval [0.0202, 0.0535], and negative for depth, −0.1222 with interval [−0.1490, −0.0957]. Its interaction is $\Gamma^{(MLP)}_{DeiT,MAE} = 0.1597$ [0.1292, 0.1889], and its modulation relative to the affine probe is $\Omega_{DeiT,MAE} = -0.0713$ [−0.0912, −0.0506]. Thus the reversal is probe-robust but decoder-modulated: its signs persist while its magnitude changes across decoder families. The paired hierarchical scene–seed bootstrap preserves the qualitative result; all ten leave-one-seed-out analyses retain Class B, and no seed is influential. Figure 11 and Table 4 summarize the result.

**Feature-perturbation sensitivity.** Adding isotropic Gaussian noise to the normalized frozen features over $\sigma \in \{0, 0.025, 0.05, 0.10, 0.20, 0.40\}$ yields the ordering DeiT < DINO < MAE within both tasks, with all paired contrasts excluding zero. The primary DeiT–MAE contrasts are −0.0750 [−0.0859, −0.0621] for semantics and −0.1786 [−0.2036, −0.1426] for depth; Table 5 reports the complete task-native AUCs and paired 95% scene/noise-bootstrap intervals. This remains an empirical sensitivity diagnostic under the declared feature scale, not an estimate of exact $L_{acc}$ or quotient geometry.

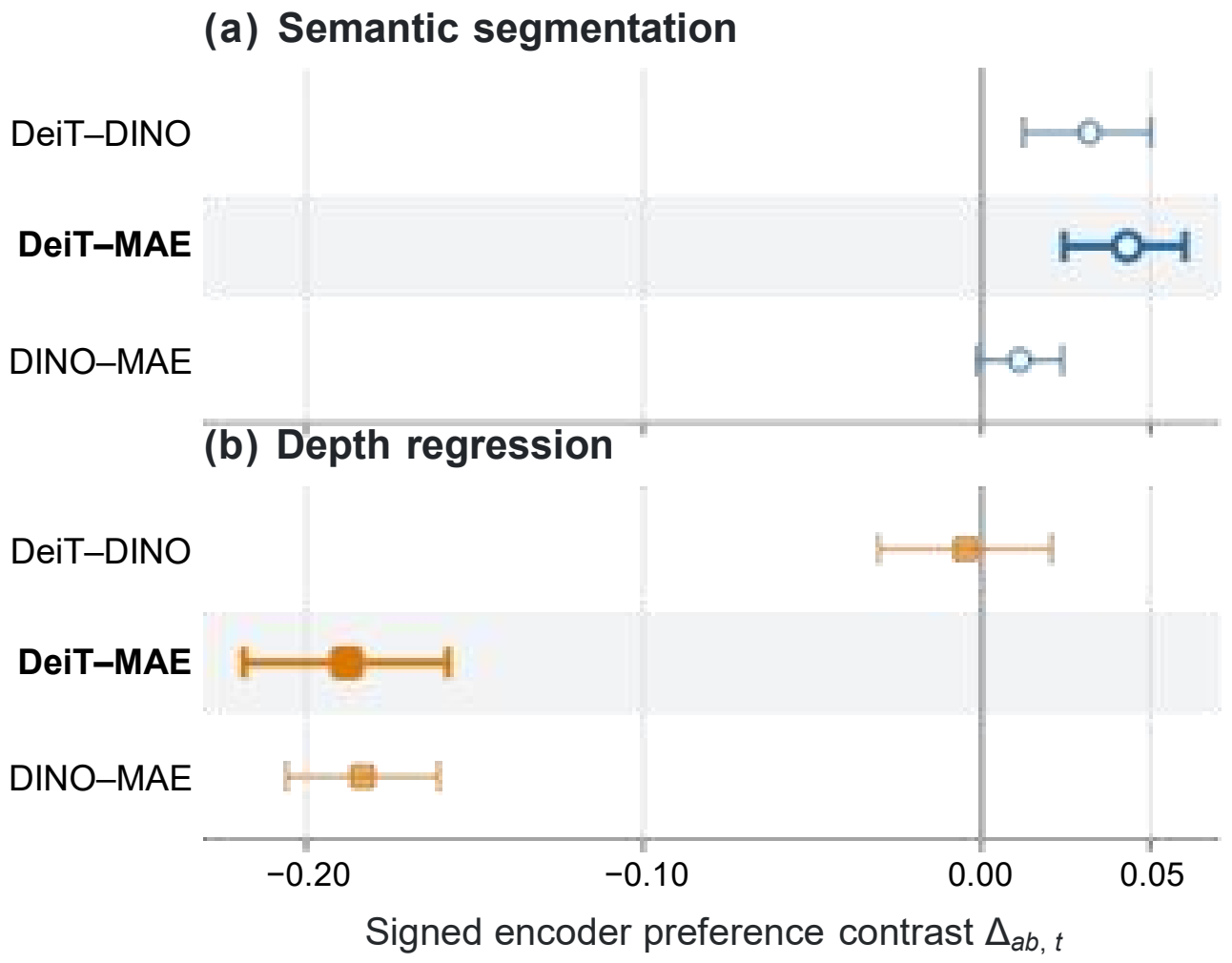

Figure 10: Encoder–task interaction on NYUv2 under frozen representations and affine probes. Signed contrasts favor the first encoder when positive; points and capped bars are estimates and paired 95% scene-bootstrap intervals. The DeiT–MAE preference reverses between semantic segmentation and depth.

Table 4: NYUv2 cross-task interactions $\Gamma_{ab} = \Delta_{ab,\mathrm{semantic}} - \Delta_{ab,\mathrm{depth}}$ under affine and MLP-128 pointwise probes. Brackets give familywise 98.33% paired scene-bootstrap intervals across the three encoder pairs.

| Encoder pair | $\Gamma_{ab}^{(affine)}$ | $\Gamma_{ab}^{(MLP-128)}$ |
|---|---|---|
| DeiT–DINO | 0.0367 [0.0005, 0.0718] | 0.0727 [0.0439, 0.1023] |
| DeiT–MAE | 0.2311 [0.1887, 0.2724] | 0.1597 [0.1234, 0.1957] |
| DINO–MAE | 0.1945 [0.1651, 0.2250] | 0.0871 [0.0583, 0.1158] |

**Finite-data affine-probe efficiency.** Training scenes were subsampled in ten paired, nested scene-level subsets at $\rho \in \{0.10, 0.25, 0.50, 0.75, 1.00\}$, while encoders, normalization, affine heads, task losses, and optimization opportunity remained fixed. All three encoders improve in their task-native metrics as data increase. The DeiT–MAE interaction remains positive from 0.2634 [0.2416, 0.2852] to 0.2307 [0.2266, 0.2347], and all pairwise interaction trajectories remain positive. Figure 12 shows the full curves, with $\rho = 1$ reproducing the full-data values. These results describe empirical finite-data efficiency for one affine probe family, not a sample-complexity theorem.

Table 5: NYUv2 relative-degradation AUC under normalized feature perturbations. Left: task-native encoder AUCs; right: paired contrasts (first minus second). Semantic and depth remain task-native quantities; brackets give paired 95% scene/noise-bootstrap intervals.

| Absolute relative-degradation AUC | | | Paired AUC contrast | | |
|---|---|---|---|---|---|
| Encoder | Semantic | Depth | Encoder pair | Semantic | Depth |
| DeiT | 0.0065 [0.0041, 0.0094] | 0.0205 [0.0171, 0.0249] | DeiT–DINO | −0.0106 [−0.0153, −0.0062] | −0.0335 [−0.0527, −0.0169] |
| DINO | 0.0171 [0.0135, 0.0212] | 0.0540 [0.0360, 0.0735] | DeiT–MAE | −0.0750 [−0.0859, −0.0621] | −0.1786 [−0.2036, −0.1426] |
| MAE | 0.0815 [0.0695, 0.0921] | 0.1991 [0.1618, 0.2236] | DINO–MAE | −0.0644 [−0.0745, −0.0519] | −0.1451 [−0.1782, −0.1157] |

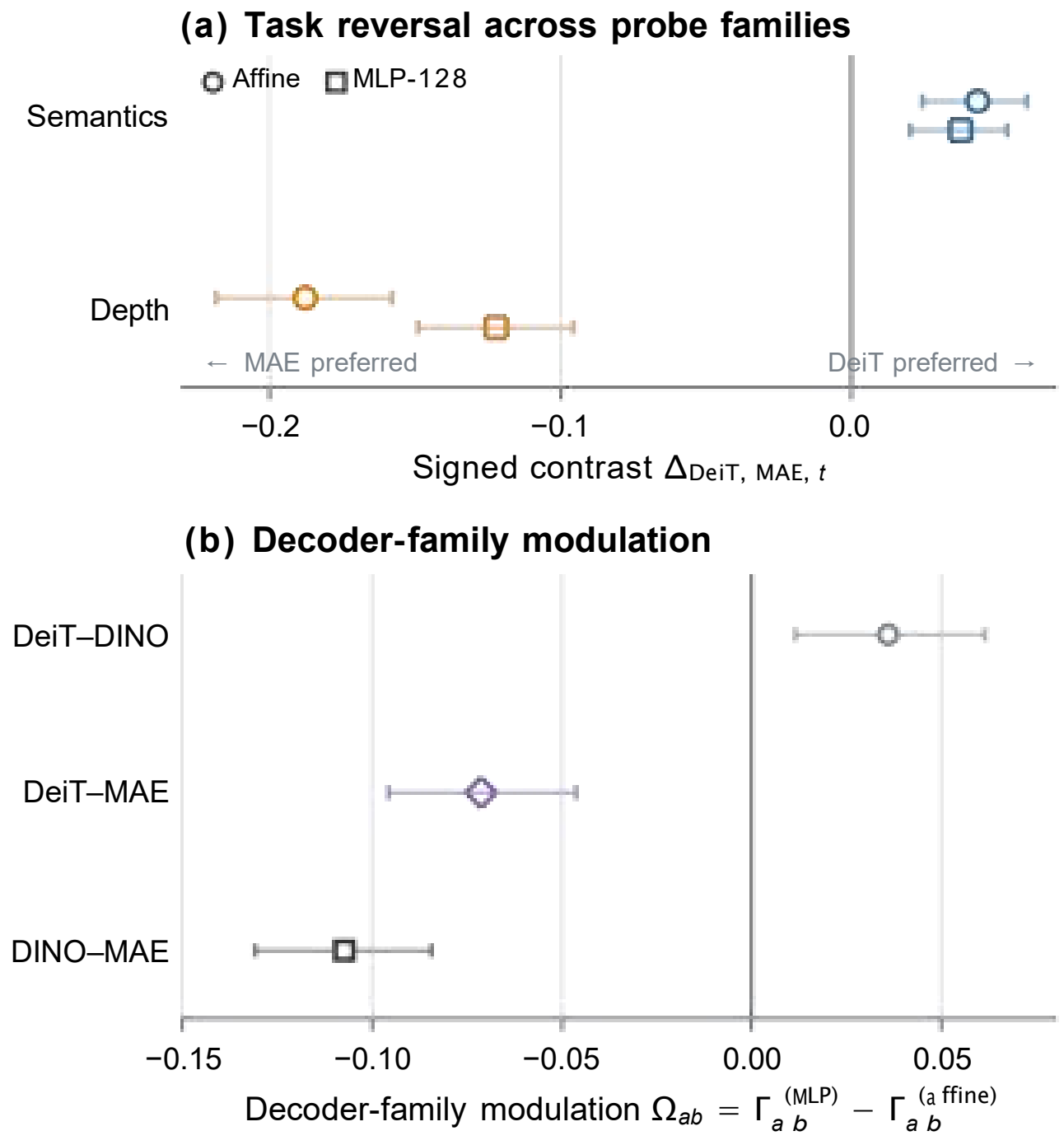

Figure 11: NYUv2 probe-family robustness. (a) Signed DeiT–MAE contrasts for affine and MLP-128 pointwise probes; positive $\Delta$ favors DeiT and negative $\Delta$ favors MAE. Intervals are paired 95% scene-bootstrap intervals. (b) Modulation $\Omega_{ab} = \Gamma_{ab}^{(\mathrm{MLP})} - \Gamma_{ab}^{(\mathrm{affine})}$ with familywise 98.33% intervals across the three encoder pairs. The reversal persists, but its magnitude is decoder-family dependent.

**Reproducibility.** All evaluations use fixed machine-readable configurations and recorded seeds, with paired scenes or subsets where appropriate; reported quantities are recomputed from machine-readable outputs. The intended code/configuration/output release also documents probe-family tuning, bootstrap and leave-one-seed-out checks, and fixed feature and checkpoint identities.

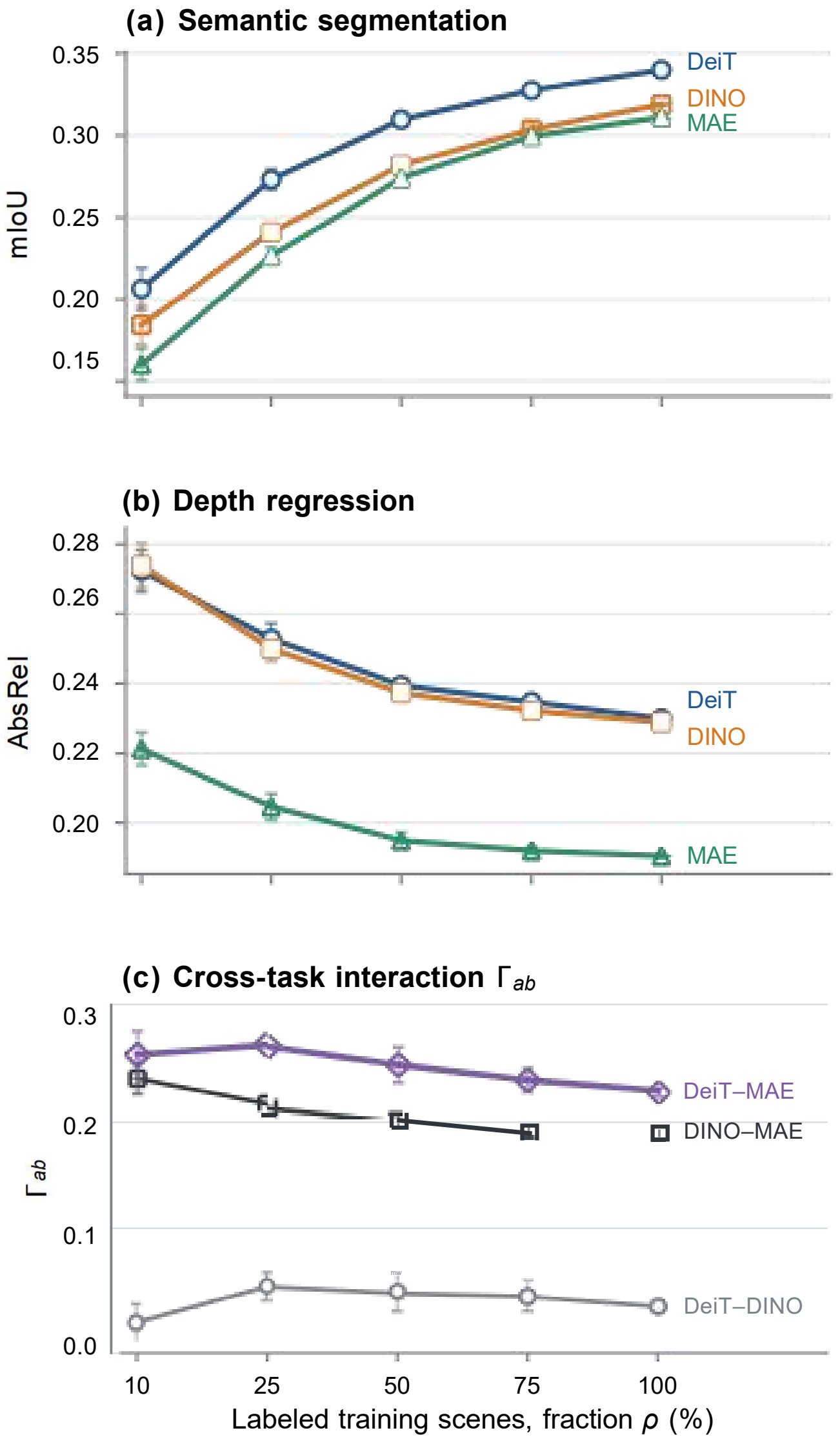

Figure 12: Finite-data affine-probe behavior on NYUv2 using ten paired, nested scene subsets. Panels (a)–(b) show mIoU and AbsRel with paired subset-seed 95% $t$ intervals; panel (c) shows $\Gamma_{ab}(\rho) = \Delta_{ab,\mathrm{semantic}}(\rho) - \Delta_{ab,\mathrm{depth}}(\rho)$. The probe family and optimization opportunity are fixed, and $\rho = 1$ reproduces the full-data study.

# 8 Discussion and Limitations

## 8.1 Implications of Task-Aware Identifiability

The first implication is that identifiability precedes learning. Before choosing an architecture or optimizer, one should ask whether the declared observation mechanism preserves the distinctions required by the task. Additional conditionally iid samples from the same fixed channel can reduce uncertainty about distinctions that survive, but they do not alter exact observation equivalence. A changed mechanism may instead refine the quotient by introducing new task-relevant information.

Second, prediction error should not be treated as a single failure mode. Structural ambiguity, weak nuisance-efficient local information, representation-channel loss, decoder accessibility, and finite learner error arise at different information levels and call for different interventions. The quotient, Fisher geometry, representation diagnostics, and layered risk decomposition provide a sequence for locating the bottleneck. The near-field case illustrates the practical consequence: two tasks supported by one sensing mechanism can occupy sharply different local-information regimes.

Third, representation quality is intrinsically task-relative. The controlled task switch and the NYUv2 encoder–task reversal provide complementary evidence that matched task-agnostic geometry or a single aggregate ranking need not determine fixed-task accessibility. The NYUv2 reversal persists under a moderately nonlinear pointwise decoder while its cross-task interaction magnitude changes across the two decoder families, a pattern consistent with decoder-regime-relative accessibility. Any such comparison depends on the task, the declared metric and scale, and the decoder regime. Accordingly, $L_{\mathrm{acc}}$ is best used as a task-conditioned stability diagnostic, not as a universal representation score.

## 8.2 Scope and Limitations

**Exact, local, and global structure.** Exact observation equivalence is an idealized boundary. Fisher geometry is a first-order description on a regular statistical image; global identifiability requires additional assumptions, and singular regimes can require higher-order analysis. Approximate equivalence and approximate task compatibility are therefore important extensions.

**Representation and metric scope.** The deterministic quotient map $\psi$ is a controlled geometric object, whereas learned encoders generally induce law-valued output experiments. Decoder accessibility depends on the task, metric, scale, and admissible decoder family; neither it nor kernel inclusion is a universal measure of neural complexity or sample efficiency. The nuisance-lift control establishes non-invariance of arbitrary ambient representatives, not the general invalidity of physical Euclidean geometry.

**Empirical coverage.** The controlled studies deliberately isolate mechanisms rather than establish universal behavior. The near-field study is synthetic and omits calibration error, mutual coupling, multipath, hardware nonlinearities, and richer nuisance structure. NYUv2 covers one indoor dataset, three ViT-B/16 encoders, two tasks, and two pointwise probe families (affine and MLP-128); it does not address spatially mixing decoders, encoder fine-tuning, broader nonlinear decoder families, or distribution-shift generalization, and the perturbation scale is tied to the declared feature normalization.

**Fixed versus adaptive sensing.** The framework diagnoses a declared observation mechanism and state-independent post-processing. It does not formulate action selection, adaptive sensor design, or closed-loop policies. The changed-mechanism control demonstrates that a new task-relevant source can alter the quotient, but it is not an active-perception algorithm.

## 8.3 Future Directions

Four directions follow directly from these boundaries: learning quotient structure when the observation mechanism is only available through data; replacing exact equivalence by controlled approximate notions; relating decoder stability to statistical and sample efficiency under explicit hypothesis-class assumptions; and extending the framework to action-dependent sensing policies that deliberately refine task-relevant observation structure.

# 9 Conclusion

Task-Aware Identifiability is a diagnostic framework that separates structural preservation of task distinctions, local statistical resolvability, representation accessibility, and downstream risk. Whether a target can be recovered is constrained first by the distinctions preserved by the observation mechanism, before representation design or learner optimization can address the distinctions that remain.

This separation localizes failure modes and supports task-relative representation evaluation across controlled models, physical sensing, and real observations. Its role is not to provide a universal representation ranking, but to determine which task-relevant distinctions survive observation, how strongly they can be resolved, and where subsequent representation or learning stages lose them.

# References

[1] R. A. Fisher, "On the mathematical foundations of theoretical statistics," *Philosophical Transactions of the Royal Society A*, vol. 222, pp. 309-368, 1922.

[2] E. L. Lehmann and G. Casella, *Theory of Point Estimation*. Springer, 1998.

[3] D. Blackwell, "Equivalent comparisons of experiments," *The Annals of Mathematical Statistics*, vol. 24, no. 2, pp. 265-272, 1953.

[4] R. E. Kalman, "A new approach to linear filtering and prediction problems," *Journal of Basic Engineering*, vol. 82, no. 1, pp. 35-45, 1960.

[5] R. Hermann and A. J. Krener, "Nonlinear controllability and observability," *IEEE Transactions on Automatic Control*, vol. 22, no. 5, pp. 728-740, 1977.

[6] T. J. Rothenberg, "Identification in parametric models," *Econometrica*, vol. 39, no. 3, pp. 577-591, 1971.

[7] S. Amari, *Differential-Geometrical Methods in Statistics*. Springer, 1985.

[8] S. Amari, *Information Geometry and Its Applications*. Springer, 2016.

[9] A. Hyvärinen and H. Morioka, "Unsupervised feature extraction by time-contrastive learning and nonlinear ICA," in *NeurIPS*, 2016.

[10] I. Khemakhem, D. Kingma, R. P. Monti, and A. Hyvärinen, "Variational autoencoders and nonlinear ICA: A unifying framework," in *AISTATS*, 2020.

[11] F. Locatello, S. Bauer, M. Lucic, G. Rätsch, S. Gelly, B. Schölkopf, and O. Bachem, "A sober look at the unsupervised learning of disentangled representations and their evaluation," *JMLR*, vol. 21, pp. 1-62, 2020.

[12] L. Li, T. J. Walsh, and M. L. Littman, "Towards a unified theory of state abstraction for MDPs," in *ISAIM*, 2006.

[13] T. Dean and R. Givan, "Model minimization in Markov decision processes," in *AAAI*, 1997.

[14] R. Bajcsy, "Active perception," *Proceedings of the IEEE*, vol. 76, no. 8, pp. 966-1005, 1988.

[15] D. V. Lindley, "On a measure of the information provided by an experiment," *The Annals of Mathematical Statistics*, vol. 27, no. 4, pp. 986-1005, 1956.

[16] K. Chaloner and I. Verdinelli, "Bayesian experimental design: A review," *Statistical Science*, vol. 10, no. 3, pp. 273-304, 1995.

[17] N. Tishby, F. C. Pereira, and W. Bialek, "The information bottleneck method," in *Proceedings of the 37th Annual Allerton Conference on Communication, Control, and Computing*, pp. 368–377, 1999.

[18] A. Achille and S. Soatto, "Emergence of invariance and disentanglement in deep representations," *Journal of Machine Learning Research*, vol. 19, no. 50, pp. 1–34, 2018.

[19] Y. Xu, S. Zhao, J. Song, R. Stewart, and S. Ermon, "A theory of usable information under computational constraints," in *International Conference on Learning Representations*, 2020.

[20] Y. Dubois, D. Kiela, D. J. Schwab, and R. Vedantam, "Learning optimal representations with the decodable information bottleneck," in *Advances in Neural Information Processing Systems*, vol. 33, 2020.

[21] N. Ay, J. Jost, H. V. Lê, and L. Schwachhöfer, "Information geometry and sufficient statistics," *Probability Theory and Related Fields*, vol. 162, no. 1–2, pp. 327–364, 2015.

[22] R. G. Baraniuk and M. B. Wakin, "Random projections of smooth manifolds," *Foundations of Computational Mathematics*, vol. 9, no. 1, pp. 51–77, 2009.

[23] A. R. Zamir, A. Sax, W. Shen, L. J. Guibas, J. Malik, and S. Savarese, "Taskonomy: Disentangling task transfer learning," in *Proceedings of the IEEE Conference on Computer Vision and Pattern Recognition*, pp. 3712–3722, 2018.

[24] F.-G. Wieland, A. L. Hauber, M. Rosenblatt, C. Tönsing, and J. Timmer, "On structural and practical identifiability," *Current Opinion in Systems Biology*, vol. 25, pp. 60–69, 2021.

[25] E. Torgersen, *Comparison of Statistical Experiments*. Cambridge University Press, 1991.

[26] L. Le Cam and G. L. Yang, *Asymptotics in Statistics: Some Basic Concepts*. Springer, 1990.

[27] D. Abel, N. Umbanhowar, K. Khetarpal, D. Arumugam, D. Precup, and M. L. Littman, "Value preserving state-action abstractions," in *Proceedings of the Twenty Third International Conference on Artificial Intelligence and Statistics*, vol. 108, pp. 1639–1650, 2020.

[28] Y. Zheng, F. Feng, Y. Li, S. Xie, K. Murphy, and K. Zhang, "From generalist to specialist representation," in *Proceedings of the 43rd International Conference on Machine Learning*, vol. 306, 2026.

[29] Y. Tian, C. Sun, B. Poole, D. Krishnan, C. Schmid, and P. Isola, "What makes for good views for contrastive learning?" in *Advances in Neural Information Processing Systems*, vol. 33, pp. 6827–6839, 2020.

[30] H. Wang, X. Guo, Z.-H. Deng, and Y. Lu, "Rethinking minimal sufficient representation in contrastive learning," in *Proceedings of the IEEE/CVF Conference on Computer Vision and Pattern Recognition*, pp. 16041–16050, 2022.

[31] Y. Dubois, T. Hashimoto, and P. Liang, "Evaluating self-supervised learning via risk decomposition," in *Proceedings of the 40th International Conference on Machine Learning*, vol. 202, pp. 8779–8820, 2023.

[32] N. Shlezinger and Y. C. Eldar, "Task-based quantization with application to MIMO receivers," *Communications in Information and Systems*, vol. 20, no. 2, pp. 131–162, 2020.

[33] Z. Wang, P. Ramezani, Y. Liu, and E. Bj¨ornson, ”Near-field localization and sensing with large-aperture arrays: From signal modeling to processing,” *IEEE Signal Processing Magazine*, vol. 42, no. 1, pp. 74–87, 2025.

[34] H. Wang, Z. Xiao, and Y. Zeng, ”Cram´er–Rao bounds for near-field sensing with extremely large-scale MIMO,” *IEEE Transactions on Signal Processing*, vol. 72, pp. 701–717, 2024.